\documentstyle[amsfonts,amssymb,graphicx,11pt]{article} %  refcheck, 
 
\newcommand{\sgn}{\mathop{\mathrm{sgn}}}

\newtheorem{theorem}{Theorem}[section]
\newtheorem{lemma}[theorem]{Lemma} 
 
\newtheorem{corollary}[theorem]{Corollary} 
\newtheorem{remark}[theorem]{Remark}

\title{\bf   
The global existence of small self-interacting scalar field propagating in the contracting universe 
}

\author{{\bf Anahit Galstian,   Karen Yagdjian}
 }

\begin{document}
\date{}
 
\maketitle
\thispagestyle{empty}
\vspace{-0.3cm}

\begin{center}
{\it School of Mathematical and Statistical Sciences,\\
University of Texas RGV,
1201 W.~University Drive, \\
Edinburg, TX 78539,
USA }
\end{center}

\thispagestyle{empty} 
\vspace{-0.3cm}

\addtocounter{section}{-1}
\renewcommand{\theequation}{\thesection.\arabic{equation}}
\setcounter{equation}{0}
\pagenumbering{arabic}
\setcounter{page}{1}
\thispagestyle{empty}

\begin{abstract}
\begin{small}
We present a condition on the self-interaction term  that guaranties  the existence of the global in time   solution of the   Cauchy problem for the semilinear Klein-Gordon equation  in the Friedmann-Lama$\hat{i}$tre-Robertson-Walker model of the contracting universe. 
For the Klein-Gordon equation with the Higgs potential we give a lower estimate  for the lifespan of  solution.


\end{small}
\end{abstract}

\section{Introduction and statement of results}
%\label{S1}

\setcounter{equation}{0}
\renewcommand{\theequation}{\thesection.\arabic{equation}}

 In the present paper we prove the global in time existence of the solutions  of the  Cauchy problem for the  semilinear \, Klein-Gordon equation in the  FLRW (Friedmann-Lama$\hat{i}  $tre-Robertson-Walker) space-time of the contracting universe for the self-interacting  scalar field.

The metric $g$ in the  FLRW space-time of the contracting universe  in the Lama$\hat{i}  $tre-Robertson  coordinates (see, e.g., \cite{Moller}) is  as follows, 
$g_{00}=  g^{00}= -  1  $, $g_{0j}= g^{0j}= 0$, $g_{ij}(x,t)=e^{-2t}    \sigma  _{ij} (x) $, 
$i,j=1,2,\ldots,n$, where $\sum_{j=1}^n\sigma  ^{ij} (x) \sigma _{jk} (x)=\delta _{ik} $,   and $\delta _{ij} $ is  Kronecker's delta. The metric $\sigma  ^{ij} (x) $ describes the time slices. 
The   covariant Klein-Gordon equation   in that space-time   in the  coordinates  is
\begin{equation}
\label{0.001}
\psi _{tt}
-  \frac{e^{2t}}{\sqrt{|\det \sigma ( x)| }} \sum_{i,j=1}^n  \frac{\partial  }{\partial x^i}\left(  \sqrt{|\det \sigma  ( x)| }  \sigma  ^{ij} (x)\frac{\partial  }{\partial x^j}  \psi \right)
- n   \psi_t +  m^2 \psi
 =
F(\psi ) \,.
\end{equation} 
It is obvious that the properties of  this equation and of its solutions are not time invertible. In fact, the equation obtained
from (\ref{0.001})  by  the time reversal   $t\rightarrow -t$ is the semilinear Klein-Gordon equation in the de~Sitter spacetime.
The Cauchy problem for the later equation is well investigated and the conditions for the existence of small data global in time solutions for some  important  metrics %$\sigma  $
 are discovered \cite{Baskin2013,Galstian-Yagdjian-NA,NARWA,Hintz-Vasy,Hintz,Nakamura,Y2017}.
%In the present paper we are interested in the Cauchy problem, which, in fact, is not equivalent to the time backward  problem for the equation with the reflected time $t\rightarrow -t$. %: 
%\begin{equation}
%\label{0.01} 
%\psi _{tt}
%-  \frac{e^{-2t}}{\sqrt{|\det \sigma ( x)| }} \sum_{i,j=1}^n  \frac{\partial  }{\partial x^i}\left(  \sqrt{|\det \sigma  ( x)| }  \sigma  ^{ij} (x)\frac{\partial  }{\partial x^j}  \psi \right)
%+ n   \psi_t +  m^2 \psi
% =
%F(\psi ) \,.
%\end{equation}
%The last equation is the semilinear Klein-Gordon equation in the de~Sitter spacetime. The equation (\ref{0.01}) is well investigated and the conditions for the existence of small data global in time solutions for some  important  metrics $\sigma  $ are discovered \cite{Baskin2013,Galstian-Yagdjian-NA,NARWA,Hintz-Vasy,Hintz,Nakamura,Y2017}.
The equation (\ref{0.001})  is a special case of the equation
\begin{eqnarray}
\label{0.1} 
&  &
  \psi  _{tt}  - e^{2 t} A(x,\partial_x) \psi -  n   \psi  _t  +  m^2  \psi  =   F(x,\psi  )\,,
 \end{eqnarray} 
where $ A(x,\partial_x) = \sum_{|\alpha |\leq 2} a_{\alpha }(x)\partial _x^\alpha $ is a  second order elliptic  partial differential operator.  We also assume that the mass $m $ can be a complex number,   $m^2 \in{\mathbb C} $. 
%If the time slices are 3D manifolds with the constant positive curvature greater than 2, then the curvature of the space-time is negative for  all positive times. The study of the Klein-Gordon equation in such space-time can be useful in catching  properties of the equation in the anti-de~Sitter space-time. The last one is   a  constant negative curvature space-time  with the  metric solving   the Einstein equations  with a negative cosmological constant in the vacuum (see \cite[Sec.~5.2]{Hawking} and \cite[p.120]{Choquet-Bruhat_book}). 

To formulate the main theorem of this paper we need some description of the nonlinear term $F$. 
 Furthermore,  we are only
concerned with the behavior of $F $ at the origin  in $\psi  $ space.  
\smallskip

\noindent {\bf Condition ($\mathcal L$).} {\it The smooth in $x$ function $F=F(x,\psi )$ is
said to be Lipschitz continuous with exponent $\alpha \geq 0 $ in the space $H_{(s)}({\mathbb R}^n)$   if there is a constant \,
$C \geq 0$ \, such that
\begin{equation}
\label{calM}
 \hspace{-0.4cm} \|  F(x,\psi  _1 )- F(x,\psi  _2 ) \|_{H_{(s)}({\mathbb R}^n)}  \leq C
\| \psi  _1 -  \psi _2   \|_{H_{(s)}({\mathbb R}^n)}
\Big( \|  \psi _1  \|^{\alpha} _{H_{(s)}({\mathbb R}^n)}
+ \|  \psi _2   \|^{\alpha} _{H_{(s)}({\mathbb R}^n)} \Big)
\end{equation}
for all \,\,  $\psi _1,\psi  _2 \in  H_{(s)}({\mathbb R}^n)$}.

\noindent
The  polynomial in $\psi  $ functions and  functions     $F(\psi  )=\pm|\psi  |^{\alpha +1}$,  $F(\psi  )=\pm|\psi  |^{\alpha } \psi  $
are important examples of the  Lipschitz continuous with exponent $\alpha > 0 $ in   the  Sobolev space $H_{(s)}({\mathbb R}^n) $, $s>n/2$, functions.   

In  what follows we   use   the metric space
\[
X({R,H_{(s)},\gamma})  := \left\{ \psi  \in C([0,\infty) ; H_{(s)}  ) \; \Big|  \;
 \parallel  \psi   \parallel _X := \sup_{t \in [0,\infty) } e^{\gamma t}  \parallel
\psi  (x ,t) \parallel _{H_{(s)}}
\le R \right\}\,,
\]
where $\gamma \in {\mathbb R}$, $R>0 $, with the metric
\[
d(\psi _1,\psi _2) := \sup_{t \in [0,\infty) }  e^{\gamma t}  \parallel  \psi _1 (x , t) - \psi _2 (x ,t) \parallel _{H_{(s)}}\,.
\]

%We denote  ${\mathcal B}^\infty $ the space of all $C^\infty ({\mathbb R}^n)$ functions with uniformly bounded derivatives of all orders. 

We define solution of the Cauchy problem  
  via corresponding integral equation 
that contains the following resolving operator  
\[
G:={\mathcal K}\circ {\mathcal EE},  
\]
where ${\mathcal EE}$ stands for the resolving operator of the evolution equation. More precisely, for the function $f(x,t) $ we define
\[
v(x,t;b):= {\mathcal EE} [f](x,t;b)\,,
\]
where the function 
$v(x,t;b)$   
is a solution to the Cauchy problem 
\begin{eqnarray}
\label{1.6} 
&   &
\partial_t^2 v - A(x,\partial_x)v =0, \quad x \in {\mathbb R}^n, \quad t \geq 0, \\
\label{2.2a}
&  &
v(x,0;b)=f(x,b)\,, \quad v_t(x,0;b)= 0\,, \quad x \in {\mathbb R}^n\,,
\end{eqnarray} 
while the integral transform ${\mathcal K}$ is given by 
\begin{eqnarray*} 
{\mathcal K}[v]  (x,t) 
&  :=  &
2   e^{\frac{n}{2}t}\int_{ 0}^{t} db
  \int_{ 0}^{ e^{t}- e^{b}} dr  \,  e^{-\frac{n}{2}b} v(x,r ;b) E(r,t; 0,b;M)  \,.
\end{eqnarray*}
Here the principal square root $M=(n^2/4-m^2)^{\frac{1}{2}}$ is the main parameter that controls estimates  and solvability of the integral equation (\ref{5.1}) that will be written below. The kernel $ E(r,t; 0,b;M) $ was introduced in \cite{JMAA_2012} and \cite{Yag_Galst_CMP} (see also (\ref{E})). Hence, 
\[
G[f]  (x,t)  
  =  
2   e^{\frac{n}{2}t}\int_{ 0}^{t} db
  \int_{ 0}^{ e^{t}- e^{b}} dr  \,  e^{-\frac{n}{2}b}\,{\mathcal EE} [f](x,r ;b) E(r,t; 0,b;M)  \,.
\] 
Thus the integral equation that corresponds to  the Cauchy problem for (\ref{0.1})   is 
\begin{eqnarray} 
\label{5.1}
\Phi (x,t)
 = 
\Phi _0(x,t) + 
G[ e^{-\Gamma \cdot }F(\cdot ,\Phi ) ] (x,t)      
\end{eqnarray}
with $\Gamma =0 $, where the function  $\Phi _0(x,t) $ is generated by the initial data. 
The main result of this paper is the following   theorem that states the existence of   global in time solution for   small initial data in Sobolev spaces.
\begin{theorem}
\label{T0.2}
Let
 $F(x,\Phi )$  be    Lipschitz continuous in the  space $H_{(s)}({\mathbb R}^n) $ with   $s>n/2\geq 1$, $F(x,0)\equiv 0$, and  $\alpha >0 $. 
Assume   that  the real part of $M $ is positive  $  \Re M >0$ and that, $M=\Re M$ if $\Re M=1/2$. If one of the following three conditions is fulfilled: 
\begin{eqnarray*} 
& (i)  &
  \frac{n}{2}+\Re M + \gamma (\alpha +1)+\Gamma >0 ,\quad   \frac{n}{2}+\max\{\frac{1}{2},\Re M \}+ \gamma  \leq  0 ,\\
& (ii) &
\frac{n}{2}+\Re M+ \gamma (\alpha +1)+\Gamma = 0 ,\quad \frac{n}{2}+\max\{\frac{1}{2},\Re M \}+ \gamma  < 0 ,\\
& (iii) &
\frac{n}{2}+\Re M + \gamma (\alpha +1)+\Gamma < 0,\quad \frac{n}{2}+\max\{\frac{1}{2},\Re M \}+\gamma \leq 0  ,\quad  \gamma  \alpha +\Gamma  \geq   0  ,
\end{eqnarray*}
then, there exists $\varepsilon _0>0 $ such that, for every $\varepsilon  < \varepsilon_0 $ and every given functions $\varphi_0 ,\varphi_1 \in H_{(s)}({\mathbb R}^n) $,  satisfying the estimate 
\[
 \| \varphi_0   \|_{H_{(s)} ({\mathbb R}^n)}
+  \|\varphi_1  \|_{ H_{(s)} ({\mathbb R}^n)} \leq \varepsilon,   
\] 
there exists a solution $\Phi \in C ([0,\infty);H_{(s)}({\mathbb R}^n))$ of the Cauchy problem  
\begin{eqnarray}
\label{0NWE}  
&  &
  \Phi _{tt} -   n   \Phi _t - e^{2 t} \Delta   \Phi  +  m^2  \Phi =  e^{-\Gamma t} F(x, \Phi )\,,\\
&  &
\label{CD0.11}
 \Phi  (x,0)= \varphi_0 (x)\, , \quad \Phi  _t(x,0)=\varphi_1 (x) \,.  
 \end{eqnarray} 
The  solution \, $ \Phi  (x ,t) $ \, belongs to the space \, $  X({2\varepsilon,H_{(s)} ({\mathbb R}^n), \gamma })  $,
that is,  
\begin{eqnarray*}  
\sup_{t \in [0,\infty)}  e^{ \gamma t}  \|\Phi  (\cdot ,t) \|_{H_{(s)} ({\mathbb R}^n)}  \leq  2\varepsilon\, .
\end{eqnarray*} 
 \end{theorem}
The theorem also includes the equations with polynomial in $\Phi  $ function $F(x,\Phi )= F(\Phi )$, with $F(0)=0 $ and the functions $F(\Phi )=\pm|\Phi |^{\alpha +1}$ and $F(\Phi )=\pm|\Phi |^{\alpha } \Phi $. 
The sharpness of the conditions on $\alpha  $, $\gamma  $, and $\Gamma  $  is an interesting open problem that will not be discussed here.

 For the given real numbers $\Gamma $,  $n$,  $\gamma  $, and the complex number $M$, define the function 
\begin{eqnarray*} 
  I(t):=  e^{ t  ( \frac{n}{2}+ \max\{\frac{1}{2},\Re M \} + \gamma )} 
\int_{ 0}^{t }   e^{-( \frac{n}{2}+\max\{\frac{1}{2},\Re M \} + \gamma (\alpha +1)+\Gamma )b}\,db  \,.
\end{eqnarray*}
\begin{theorem}
\label{T0.2b}
%$\mbox{\rm (LfS)}$ 
If $\Re M > 0$, $ \gamma  \leq -(\frac{n}{2}+\max\{\frac{1}{2},\Re M \} ) $, and none of the  conditions (i)-(iii) is fulfilled, then  
the lifespan $T_{ls}$ of the solution of (\ref{0NWE})-(\ref{CD0.11}) can be estimated from below as follows
\[  
  T_{ls}    
   \geq   {\cal I} \left(C (M,n,\alpha ,\gamma, \Gamma  )    \left( \| \varphi_0   \|_{ H_{(s)} ({\mathbb R}^n)}
+  \|\varphi_1  \|_{H_{(s)} ({\mathbb R}^n)} \right)  ^{-\alpha }\right)  
\]
with some constant $C (M,n,\alpha ,\gamma, \Gamma  ) $ for  sufficiently small $\| \varphi_0   \|_{ H_{(s)} ({\mathbb R}^n)}
+  \|\varphi_1  \|_{ H_{(s)} ({\mathbb R}^n)} $. Here  $\cal I$ is the function inverse   to $I=I(t)$. 
 \end{theorem}

 In particular, Theorem~\ref{T0.2b} states  an estimate for the lifespan $T_{ls}$ of the solution of the  Higgs boson equation with the Higgs potential  in the contracting  universe; which reads 
\[
\psi _{tt} - e^{2t} \Delta \psi - n\psi_t  =  \mu^2  \psi  - \lambda \psi  ^3 \,.
\] 
Here    $\lambda >0 $ and $\mu >0 $. For instance, if $n=3$, $\alpha =2$, $\Gamma =0$, $\gamma =-3/2 $, $\mu ^2=  7/4$, that is $m^2=-7/4$ and $M=2 $, then 
\[  
  T_{ls}    
   \geq  - \frac{2}{3}\ln \left( \| \varphi_0   \|_{ H_{(s)} ({\mathbb R}^n)}
+  \|\varphi_1  \|_{H_{(s)} ({\mathbb R}^n)} \right)+C\,.   
\]

\noindent
{\bf Example.} To discuss the conditions of Theorem~\ref{T0.2} we consider  equation relevant to the physically interesting case of  $n=3 $ and $\alpha =2$.  
Then for the real $M$ the conditions (i)-(iii) read as follows.
%\begin{eqnarray*} 
%& (i)  &
%  \frac{3}{2}+\Re M + 3 \gamma  +\Gamma >0 ,\quad   \frac{3}{2}+\max\{\frac{1}{2},\Re M \}+ \gamma  \leq  0 ,\\
%& (ii) &
%\frac{3}{2}+\Re M+ 3\gamma +\Gamma = 0 ,\quad \frac{3}{2}+\max\{\frac{1}{2},\Re M \}+ \gamma  < 0 ,\\
%& (iii) &
%\frac{3}{2}+\Re M + 3\gamma  +\Gamma < 0,\quad \frac{3}{2}+\max\{\frac{1}{2},\Re M \}+\gamma \leq 0  ,\quad  2\gamma   +\Gamma  \geq   0  .
%\end{eqnarray*}
%If $M=1/2 $, then
%\begin{eqnarray*} 
%& (i_{1/2})  &
%  2 + 3 \gamma  +\Gamma >0 ,\quad   2+ \gamma  \leq  0 ,\\
%& (ii_{1/2}) &
%2+ 3\gamma +\Gamma = 0 ,\quad 2+ \gamma  < 0 ,\\
%& (iii_{1/2}) &
%2+ 3\gamma  +\Gamma < 0,\quad 2+\gamma \leq 0  ,\quad  2\gamma   +\Gamma  \geq   0  .
%\end{eqnarray*}
%\begin{figure}[ht]
%\caption{ \small Feasible domain for $ \gamma ,\Gamma $ if $M=1/2$, $n=3$ and $\alpha =2$}
%\centering{} 
%\begin{tabular}{cc}
%{\tiny \bf (a)  Case i of $ M=1/2$, $-10<\gamma <-2 $, $4< \Gamma< 6 $}  & {\tiny \bf (b) Case iii of $  M=1/2$, $-10<\gamma <-2 $, $4< \Gamma< 6 $ }\tabularnewline 
%  \includegraphics[height=4cm]{Mequal12i.eps} &  \includegraphics[height=4cm]{Mequal12iii.eps} \tabularnewline   
%\end{tabular}
%\end{figure}
If $M\leq 1/2 $, then 
\begin{eqnarray*} 
& (i_{M\leq 1/2})  &
  \frac{3}{2}+  M + 3 \gamma  +\Gamma >0 ,\quad   2+ \gamma  \leq  0 ,\\
& (ii_{M \leq 1/2}) &
\frac{3}{2}+  M+ 3\gamma +\Gamma = 0 ,\quad  2+ \gamma  < 0 ,\\
& (iii_{M \leq 1/2}) &
\frac{3}{2}+  M + 3\gamma  +\Gamma < 0,\quad    2+ \gamma  \leq  0 ,\quad    2\gamma   +\Gamma  \geq   0  .
\end{eqnarray*}
If $M>1/2 $, then 
\begin{eqnarray*} 
& (i_{M>1/2})  &
  \frac{3}{2}+ M + 3 \gamma  +\Gamma >0 ,\quad   \frac{3}{2}+M+ \gamma  \leq  0 ,\\
& (ii_{M>1/2}) &
\frac{3}{2}+  M+ 3\gamma +\Gamma = 0 ,\quad \frac{3}{2}+M+ \gamma  < 0 ,\\
& (iii_{M>1/2}) &
\frac{3}{2}+  M + 3\gamma  +\Gamma < 0,\quad \frac{3}{2}+M+\gamma \leq 0  ,\quad  2\gamma   +\Gamma  \geq   0  .
\end{eqnarray*}
\begin{figure}[ht]
\caption{\small Feasible domain for $M,\gamma ,\Gamma $ if $n=3$ and $\alpha =2$}
\centering{} 
\begin{tabular}{cc}
{\tiny \bf (a)  Case i of $0<M<1/2$, $-3<\gamma <-2 $, $4< \Gamma<6 $}  & {\tiny \bf (b) Case iii of $0< M<1/2$, $-3<\gamma <-2 $, $4< \Gamma< 6 $ }\tabularnewline 
  \includegraphics[height=5cm]{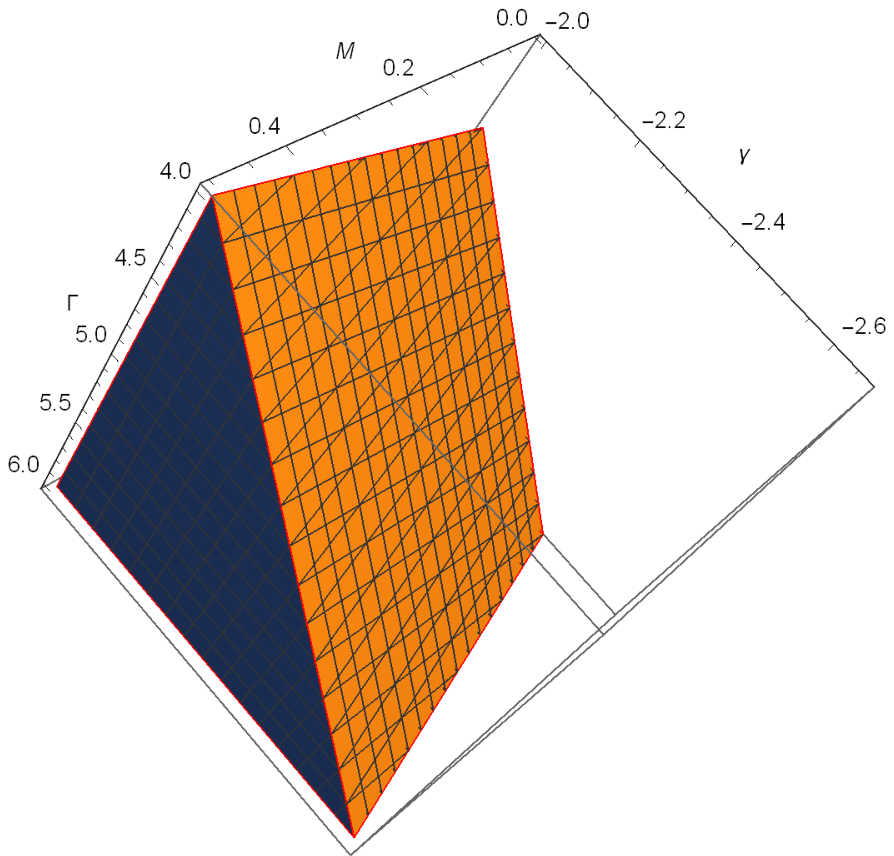} &  \includegraphics[height=5cm]{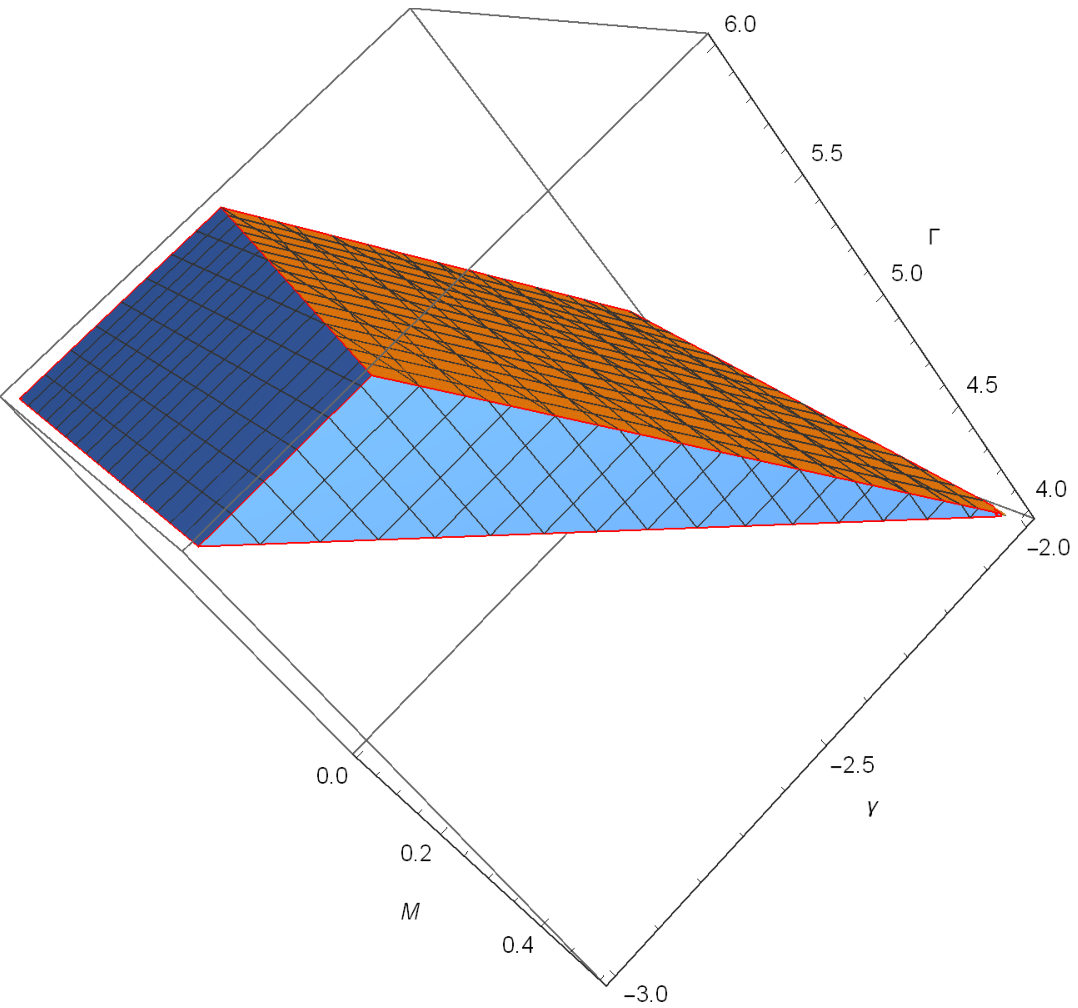} \tabularnewline  
{\tiny  \bf (c) Case i of $1/2<M<3/2$, $-3<\gamma <-2 $, $4< \Gamma< 6 $}  & {\tiny \bf (d)  Case iii of $1/2<M<3/2$, $-3<\gamma <-2 $, $4< \Gamma< 6 $} \tabularnewline
  \includegraphics[height=5cm]{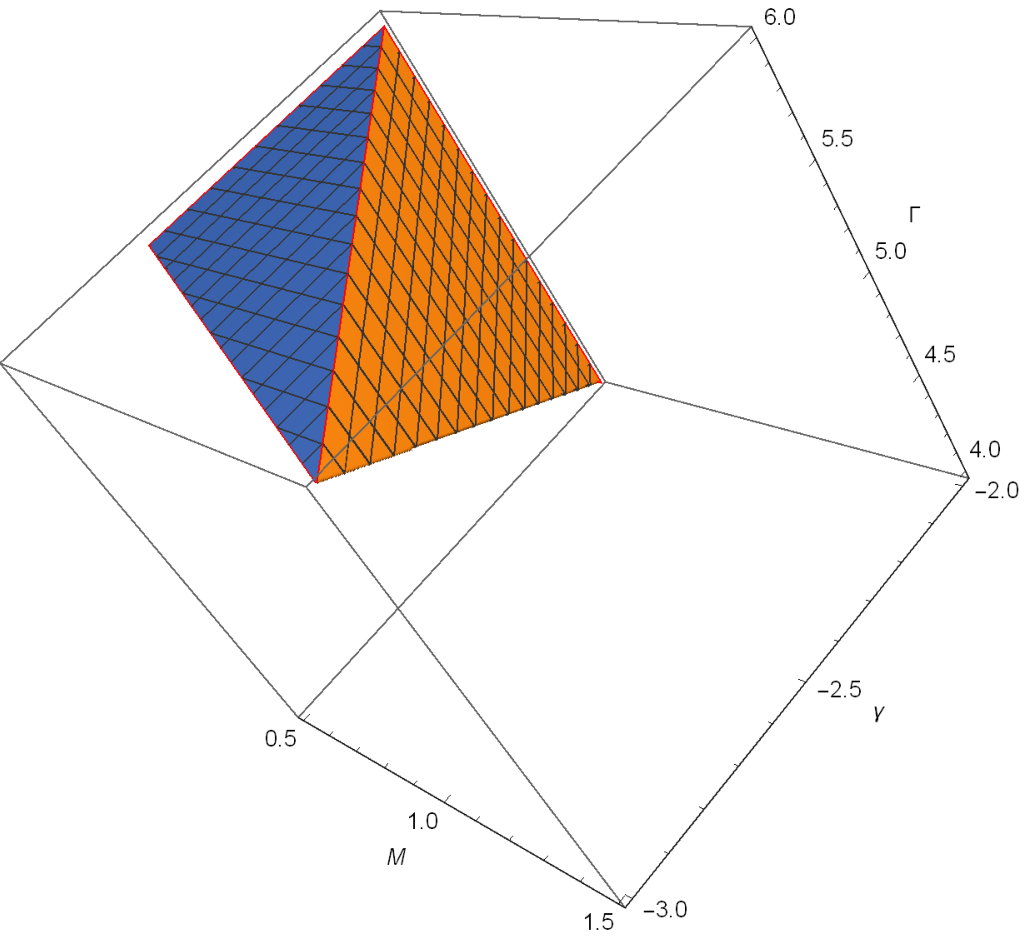} & \includegraphics[height=5cm]{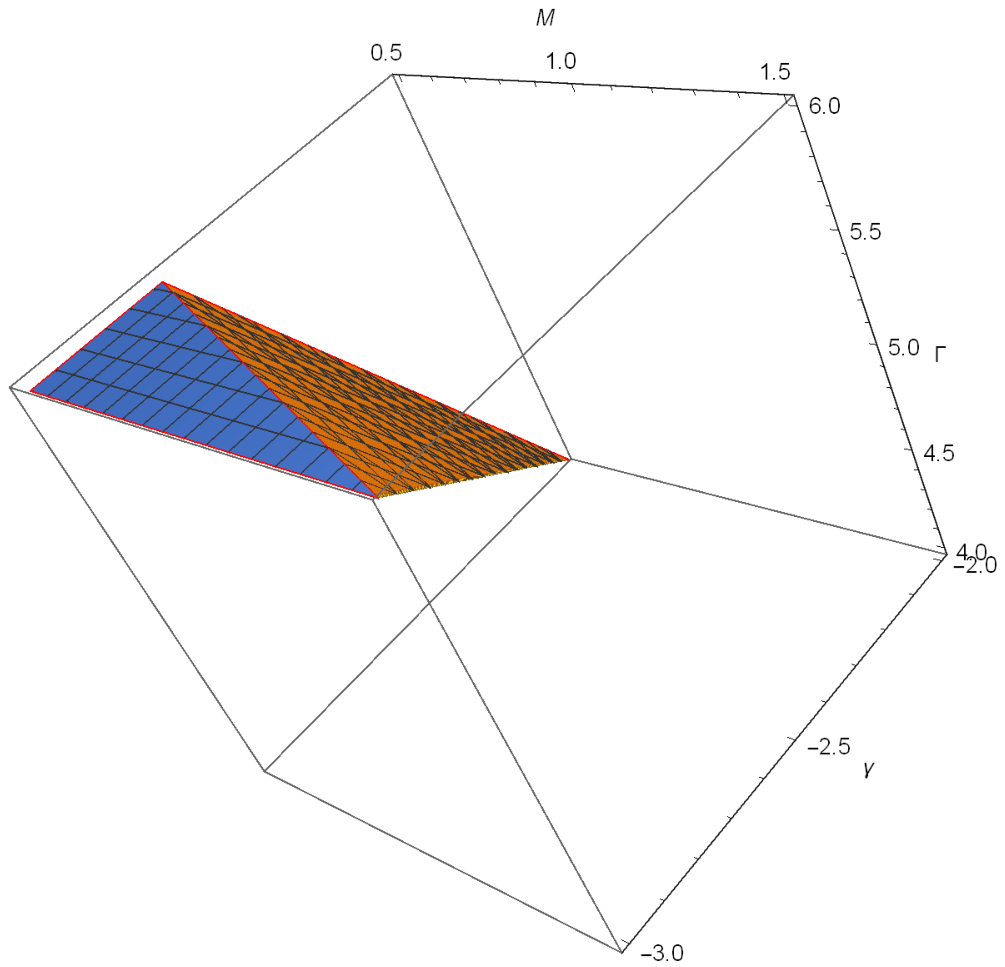}  
\end{tabular}
\end{figure}

In order to prove  Theorem~\ref{T0.2} and Theorem~\ref{T0.2b}  we establish in Section~\ref{S30}   estimates in the Besov spaces $ B^{s,q}_{p}$ for the linear equation obtained from (\ref{0.1}) although estimates in the Sobolev spaces are sufficient for our goal. In fact, the proofs of the  estimates for these two cases are identical. For the linear equation without a source term,  these estimates for the large time $t$ imply the limitation for the rate of growth as follows: 
\[  
\| \Phi  (x,t) \| _{X'}  \nonumber   
   \leq    
  C e^{(\frac{n}{2}+a+ \max \{\frac{1}{2}, \Re M\})t  } \|\varphi _0 \|_{X}    
+ C e^{(\frac{n}{2}+a+ \Re M)t}  \|\varphi _1 \|_{X}  \,,
\]   
%\[  
%\| \Phi  (x,t) \| _{X'}  \nonumber   
 %  \leq    
%\|\varphi _0 \|_{X} \Bigg( C e^{(\frac{n}{2}+a)t}   
%+   e ^{ \Re M t} 
%\cases{   e^{  \frac{1}{2} t} \quad \mbox{\rm if} \quad \Re M<1/2\,,\cr
%  e^{ \Re M t} \quad \mbox{\rm if}  \quad \Re M\geq 1/2}  \Bigg)  
%+ \|\varphi _1 \|_{X}C e^{(\frac{n}{2}+a+ \Re M)t}    \,,
%\]   
where if $X=B^{s,q}_{p}$, then $X'=B^{s',q}_{p'}$,  $a:=s-s'-2n\left( 1/p-1/2\right) $, $1/p+1/p'=1$, while $X'=L^{p'}$ if $X=L^p $.
In the case of Sobolev spaces  $X=X'=H_{(s)}({\mathbb R}^n)$  and $p=2$. 
%In fact, due to \cite{Brenner}  
%(see Theorem~\ref{TBrenner} of the next section) such estimate is true also for the operators with $A=A(x,D)$ being  a second order negative elliptic differential operator with real-valued $C^\infty$-coefficients such that 
%$ A(x,D) =A(\infty,D)$ for $|x|$ large enough. 
The integral transform ${\mathcal K}$ allows us to avoid consideration in the phase space   and to apply immediately the  well-known decay estimates  for the solution of  the wave equation (operator ${\mathcal EE} $) (see, e.g., \cite{Brenner,Brenner-Kumlin,Pecher}).  

 Ebert and Nunes  do Nascimento    \cite{E-N}  have studied  the long time behavior of the energy of solutions
for a class of linear equations with time-dependent mass and speed of propagation.
They introduced a classification of the potential term, which clarifies
whether the solution behaves like the solution to the wave equation or the Klein-Gordon equation. For the equation 
\[
u_{tt} - e^{2t} \Delta u +m^2  u    =   |u|^p ,
\]
with $n \leq 4$, $m>0$, $2\leq p\leq \frac{n}{[n-2]_+} $  they established the existence of energy class solution for small data.  Their proof is based on   splitting the phase space into pseudo-differential and hyperbolic zones. That method   of zones was invented for the  hyperbolic operators with multiple characteristics (see \cite{Yagdjian_book} and references therin) and then modified and successfully used in studying the   large time  behavior of the solutions   (see \cite{E-R-book,Galstian,Hirosawa-Wirth,Reissig-Yagdjian-MN,Wirth2004,Wirth2007},  and references therein).

Hirosawa and Nunes do Nascimento~\cite{Hirosawa-Nunes} have proved some energy estimates   for Klein-Gordon equations with
time-dependent potential:
\[
u_{tt} -  \Delta u +m(t)  u    = 0\,,
\]
where 
\[
m(t)=\frac{\mu ^2}{(1+t)^2}+\delta (t), \quad 0<\mu <\frac{1}{2}, \quad |\delta (t) | \lesssim (1+t)^{-2\beta }, \quad \beta <1\,,
\]
under some conditions on the possible oscillations of $\delta (t) $. Here, $m(t) $ is not necessarily positive function. 

The present paper is organized as follows. In Section~\ref{S1} we describe 
 the  integral transform from \cite{MN} and the  representations   generated by that transform  
for the solutions of the Cauchy problem for the linear equation.   In Section~\ref{S2} some estimates for the kernels of those integral transformers are derived.  Then, in Section~\ref{S30} we quote  the $ B^{s,q}_{p}-B^{s',q}_{p'}$-estimates from \cite{Brenner} 
for the second order hyperbolic operator and  using integral transform from Section~\ref{S1}  we demonstrate   how  these estimates can be pushed forward 
to the source free equation with the  exponentially increasing in time coefficient.  In Section~\ref{S2b}  we obtain similar estimates for the equation with the source term. 
 Section~\ref{S3} is devoted to the solvability of the associated integral equation while Section~\ref{S4} completes the proof of 
 Theorems~\ref{T0.2}-\ref{T0.2b}.

\section{Resolving operators for the linear equation}
\label{S1}

\setcounter{equation}{0}
\renewcommand{\theequation}{\thesection.\arabic{equation}}

The following partial Liouville transform (change of unknown function) 
\[
u= e^{-\frac{n}{2}t}\Phi  \,,\qquad \Phi  = e^{\frac{n}{2}t}u\,,
\]
eliminates the term with time derivative of the equation (\ref{0NWE}). We obtain 
%\[
%u _{tt} -   e^{2 t} A(x,\partial_x) u   +  \left( m^2 -\frac{n^2}{4} \right) u =  e^{-(\Gamma +\frac{n}{2})t} F(e^{\frac{n}{2}t}u )\,,
% \]
% which can be written as follows 
\begin{equation}
\label{K_G_Higgs}
u _{tt} -   e^{2 t} A(x,\partial_x) u   - M^2   u =  e^{-(\Gamma +\frac{n}{2})t} F(e^{\frac{n}{2}t}u )\,, 
\end{equation} 
 where
 \[
 M=\left( \frac{n^2}{4} - m^2\right)^{\frac{1}{2}}\,.  
 \]
In this section we consider the linear part of the equation (\ref{K_G_Higgs})  
with $M \in {\mathbb C} $. The  equation (\ref{K_G_Higgs}) covers two important cases. The first one is the Higgs boson equation, which has $F(\phi )=-\lambda \phi ^3 $ 
and $M^2= \frac{n^2}{4}+ \mu^2    $ with $\lambda >0 $, $\mu >0 $, and   $n=3$. 
This includes also equation of {\it tachyonic scalar fields} living on the   de~Sitter universe. (See, e.g. \cite{Bros-2,Epstein-Moschella}.)  The second case is the case  of the small physical  mass (the light scalar field), that is $0 \leq m  \le \frac{n }{2}$. %For the last case  $ M = \sqrt{\frac{n^2}{4}-m^2}$. 

We introduce  the kernel functions $E(x,t;x_0,t_0;M) $, $K_0(z,t;M)   $,    and $K_1(z,t;M) $    (see also \cite{JMAA_2012} and \cite{Yag_Galst_CMP}). 
 First, for $M \in {\mathbb C} $  we define the function 
\begin{eqnarray}
\label{E}  
E(x,t;x_0,t_0;M)
 &  = & 
 4 ^{-M}  e^{ -M(t_0+t) } \Big((e^{t }+e^{t_0})^2 - (x - x_0)^2\Big)^{-\frac{1}{2}+M    } \\
 &  &
 \times 
F\Big(\frac{1}{2}-M   ,\frac{1}{2}-M  ;1; 
\frac{ ( e^{t}-e^{t_0 })^2 -(x- x_0 )^2 }{( e^{t}+e^{t_0 })^2 -(x- x_0 )^2 } \Big) . \nonumber 
\end{eqnarray} 
Next 
we define also the kernels  $K_0(z,t;M)   $    and $K_1(z,t;M) $ by
\begin{eqnarray*}
K_0(z,t;M)
& := & 
- \left[  \frac{\partial }{\partial b}   E(z,t;0,b;M) \right]_{b=0} \\
&  = &   
-4 ^ {-M}  e^{- t M}\big((1+e^{t })^2 - z^2\big)^{  M    } \frac{1}{ [( e^{t}-1)^2 -  z^2]\sqrt{(e^{t }+ 1)^2 - z^2} }\\
&   &
\times  \Bigg[  \big(  e^{t} -1 +M(e^{ 2t} -      1 -  z^2) \big) 
F \Big(\frac{1}{2}-M   ,\frac{1}{2}-M  ;1; \frac{ ( e^{t}-1)^2 -z^2 }{( e^{t }+1)^2 -z^2 }\Big) \\
&  &
\hspace{1cm}  +   \big( 1-e^{2 t}+  z^2 \big)\Big( \frac{1}{2}+M\Big)
F \Big(-\frac{1}{2}-M   ,\frac{1}{2}-M  ;1; \frac{ ( e^{t}-1)^2 -z^2 }{( e^{t }+1)^2 -z^2 }\Big) \Bigg]
\end{eqnarray*} 
and $K_1(z,t;M)   :=  
  E(z ,t;0,0;M) $, that is, 
\begin{eqnarray*} 
K_1(z,t;M)  
& = &
  4 ^{-M} e^{- Mt }  \big((e^{t }+1)^2 -   z  ^2\big)^{-\frac{1}{2}+M    } \\
  &  &
  \times 
F\left(\frac{1}{2}-M   ,\frac{1}{2}-M  ;1; 
\frac{ ( e^{t}-1)^2 -z^2 }{( e^{t }+1)^2 -z^2 } \right), \, 0\leq z\leq  e^{t}-1, 
 \end{eqnarray*}  
respectively.  
For the solution  $\Phi $ of the the Cauchy problem  
\[
  \Phi _{tt} -  n   \Phi _t - e^{2 t} A(x,\partial_x) \Phi  + m^2\Phi =  f ,\quad \Phi  (x,0)= \varphi _0(x)  , \quad \Phi  _t(x,0)=\varphi _1(x), 
\]
according to \cite{MN} we obtain 
\begin{eqnarray}
\label{Phy}
&  &
\Phi  (x,t) \\
 & =  &
2   e^{\frac{n}{2}t}\int_{ 0}^{t} db
  \int_{ 0}^{ e^{t}- e^{b}} dr  \,  e^{-\frac{n}{2}b} v_f(x,r ;b) E(r,t; 0,b;M)   \nonumber \\
&  &
+ e^{\frac{n-1}{2}t} v_{\varphi_0}  (x, \phi (t))
+ \,  e^{\frac{n}{2}t}\int_{ 0}^{1} v_{\varphi_0}  (x, \phi (t)s)\big(2  K_0(\phi (t)s,t;M)- nK_1(\phi (t)s,t;M)\big)\phi (t)\,  ds  \nonumber \\
& &
+\, 2e^{\frac{n}{2}t}\int_{0}^1   v_{\varphi _1 } (x, \phi (t) s) 
  K_1(\phi (t)s,t;M) \phi (t)\, ds
, \quad x \in {\mathbb R}^n, \,\, t>0\, , \nonumber 
\end{eqnarray}
where the function 
$v_f(x,t;b)$   
is a solution to the Cauchy problem  (\ref{1.6})-(\ref{2.2a}),  
while $\phi (t):= e^{t} - 1$. 
 Here, for $\varphi \in C_0^\infty ({\mathbb R}^n)$ and for $x \in {\mathbb R}^n$, 
the function $v_\varphi  (x, \phi (t) s)$  coincides with the value $v(x, \phi (t) s) $ 
of the solution $v(x,t)$ of the Cauchy problem for the equation (\ref{1.6}) with the first initial datum $\varphi  (x) $,  while the second datum is zero.

\section{Some estimates for the kernel functions $\bf K_0$ and $\bf K_1$}
\label{S2}
\setcounter{equation}{0}
\renewcommand{\theequation}{\thesection.\arabic{equation}}

\begin{lemma}
\label{L2.3}
Let $  a>-1 $, $\Re M>0$,  and $\phi (t)=e^{t} -1$. Then 
\begin{eqnarray*} 
 \int_{ 0}^{1} \phi (t)^{a} s^{a}
\big|  K_1(\phi (t)s,t;M)\big|   \phi (t)\,  ds  
& \leq &
 C_M  e^{-\Re Mt }(e^{t }-1)^{a+1} (e^{t }+1)^{2 \Re M-1}  \quad \mbox{\rm for all} \quad t>0\,.
\end{eqnarray*}  
In particular,
\begin{eqnarray*} 
 \int_{ 0}^{1} \phi (t)^{a} s^{a}
\big|  K_1(\phi (t)s,t;M)\big|   \phi (t)\,  ds  
& \leq &
 C_{M,a}    e^{(\Re M+a)t}  \quad \mbox{  for all}  \quad t \in [1,\infty) \,,\\ 
 \int_{ 0}^{1} \phi (t)^{a} s^{a}
\big|  K_1(\phi (t)s,t;M)\big|   \phi (t)\,  ds  
& \leq &
 C_{M,a}  t^{a+1}  \quad \mbox{  for all}  \quad t \in (0,1)\,.
\end{eqnarray*}
\end{lemma}
\medskip

\noindent
{\bf Proof.} By the definition of the kernel $K_1$, we obtain
\begin{eqnarray*} 
&  &
 \int_{ 0}^{1} \phi (t)^{a} s^{a}
\big|  K_1(\phi (t)s,t;M)\big|   \phi (t)\,  ds   
   =    
 \int_{ 0}^{e^{t}-1} r^{a} 
\big|  K_1(r,t;M)\big|    \,  dr \\
& \leq &  
   4 ^{-\Re M} e^{ -\Re Mt }  \int_{ 0}^{e^{t}-1} r^{a} 
  \big((e^{t }+1)^2 -   r  ^2\big)^{-\frac{1}{2}+\Re M    }\left|F\left(\frac{1}{2}-M   ,\frac{1}{2}-M  ;1; 
\frac{ ( e^{t}-1)^2 -z^2 }{( e^{t }+1)^2 -z^2 } \right)  \right|   \,  dr\,.  
\end{eqnarray*}
On the other hand, for $\Re M>0$ we have (see pages 8,9~\cite{Y2017}) 
\begin{eqnarray*} 
 \int_{ 0}^{1} \phi (t)^{a} s^{a}
\big|  K_1(\phi (t)s,t;M)\big|   \phi (t)\,  ds 
& \leq &
 C_M  e^{-\Re Mt  } \int_{ 0}^{e^{t }-1}  y^{ a} 
  \big((e^{t }+1)^2 -   y  ^2\big)^{-\frac{1}{2}+\Re M    }     \,   dy \,.
\end{eqnarray*}
 and again due to   pages 8,9~\cite{Y2017} for $M>0$ we have
\begin{eqnarray} 
\label{1.10}
&  &
\int_{ 0}^{z-1}  y^{a} 
  \big((z+1)^2 -   y  ^2\big)^{-\frac{1}{2}+M    }     \,   dy  \\
& = &
\frac{1}{1+a}(z-1)^{1+a} (z+1)^{2 M-1} F\left(\frac{1+a}{2},\frac{1}{2}-M;\frac{3+a}{2};\frac{(z-1)^2}{(z+1)^2}\right) , \nonumber 
 \end{eqnarray}  
where $a>-1$ and $z\geq 1$. Hence, for $\Re M>0$ we have
\begin{eqnarray*} 
\int_{ 0}^{1} \phi (t)^{a} s^{a}
\big|  K_1(\phi (t)s,t;M)\big|   \phi (t)\,  ds   
&  \leq &
C_M  e^{-\Re Mt }(e^{t }-1)^{a+1} (e^{t }+1)^{2 \Re M-1}  \quad \mbox{\rm for all} \quad t>0\,.
\end{eqnarray*}
Thus the lemma is proved. \hfill $\square$

\begin{lemma}
\label{L1.2}
Let $ a>-1$,  $\Re M>0$, and  $\phi (t)=e^{t}- 1$. Then  
\begin{eqnarray*}
&  &
 \int_{ 0}^{1} \phi (t)^{a} s^{a}
\big|  K_0(\phi (t)s,t;M)\big|   \phi (t)\,  ds  \nonumber \\   
& \leq  &
  C_{M,a} (e^t-1)^{a+1} (e^t+1)^{  \Re M-1}   
\cases{ 1 \quad \mbox{\rm if } \quad   \Re \, M>1/2 \cr  
t^{\sgn|\Im M|}+ e^{(\frac{1}{2}-\Re M)t  }   \quad \mbox{\rm if } \quad \Re \,M  \leq  1/2 }  \   
\end{eqnarray*}
for all   $t>0$. In particular, for all $t \in [1,\infty) $ 
\begin{eqnarray*} 
 \int_{ 0}^{1} \phi (t)^{a} s^{a}
\big|  K_0(\phi (t)s,t;M)\big|   \phi (t)\,  ds  
& \leq & 
  C_{M,a} e^{t(a+\Re M)}   
\cases{ 1 \quad \mbox{\rm if } \quad   \Re \, M>1/2\,, \cr  
t^{\sgn|\Im M|}+ e^{(\frac{1}{2}-\Re M)t  }   \quad \mbox{\rm if } \,\, \Re \,M  \leq  1/2  ,}
\end{eqnarray*} 
while
\begin{eqnarray*}
 \int_{ 0}^{1} \phi (t)^{a} s^{a}
\big|  K_0(\phi (t)s,t;M)\big|   \phi (t)\,  ds  
& \leq & C_{M,a}  
 t^{a+1} \quad \mbox{\rm for all} \quad t  \in(0,1)\,.
\end{eqnarray*} 
\end{lemma}
\medskip

\noindent
{\bf Proof.} It is evident that 
\begin{eqnarray*} 
&  &
 \int_{ 0}^{1} \phi (t)^{a} s^{a}
\big|  K_0(\phi (t)s,t;M)\big|   \phi (t)\,  ds   
   =    
 \int_{ 0}^{e^{t}-1} y^{a} 
\big|  K_0(y,t;M)\big|    \,  dy \,.  
\end{eqnarray*}
By definition of the kernel $K_0$  we obtain with  $z=e^t$ 
\begin{eqnarray}
\label{2.2new} 
&  & 
 \int_{ 0}^{1} \phi (t)^{a} s^{a}
\big|  K_0(\phi (t)s,t;M)\big|   \phi (t)\,  ds \\
& \leq &
4 ^ {-\Re M}  z^{ - \Re M}  \int_{ 0}^{z-1 }  y^{a}
  \big((1+z)^2 - y^2\big)^{  \Re M    } \frac{1}{ [( z-1)^2 -  y^2]\sqrt{(z+ 1)^2 - y^2} } \nonumber \\
&   &
\times  \Bigg|\Bigg[  \big(  z -1 +M(z^{ 2 } -      1 -  y^2) \big) 
F \Big(\frac{1}{2}-M   ,\frac{1}{2}-M  ;1; \frac{ ( z-1)^2 -y^2 }{( z+1)^2 -y^2 }\Big)  \nonumber \\
&  &
\hspace{1cm}  +   \big( 1-z^{2  }+  y^2 \big)\Big( \frac{1}{2}+M\Big)
F \Big(-\frac{1}{2}-M   ,\frac{1}{2}-M  ;1; \frac{ ( z-1)^2 -y^2 }{( z+1)^2 -y^2 }\Big) \Bigg]\Bigg|    \,  d  y  \,. \nonumber 
\end{eqnarray}
Then we follow arguments used in the proof of     Proposition 1.6~\cite{Y2017}.
For the large number $N$ if $z\in(1,N)$, then
\[
\frac{ (z- 1)^2 - y ^2 }{(z+ 1)^2 - y ^2 } \leq \frac{ (z- 1)^2   }{(z+ 1)^2  } \leq \frac{ (N- 1)^2   }{(N+ 1)^2  } <1 \quad \mbox{\rm for all} \quad y \in (0,z-1)\,.
\]
Here and henceforth, if $A$ and $B$ are two non-negative quantities, we use $A \lesssim  B$ to denote the statement that $A\leq CB $ for some absolute constant $C>0$. For these values of $z$  and $\Re M\geq 0$ we consider the function
\begin{eqnarray*} 
&  &
 \frac{1}{ [( z-1)^2 -  y^2] } \Bigg|\Bigg[  \big(  z -1 +M(z^{ 2 } -      1 -  y^2) \big) 
F \Big(\frac{1}{2}-M   ,\frac{1}{2}-M  ;1; \frac{ ( z-1)^2 -y^2 }{( z+1)^2 -y^2 }\Big) \\
&  &
\hspace{1cm}  +   \big( 1-z^{2  }+  y^2 \big)\Big( \frac{1}{2}+M\Big)
F \Big(-\frac{1}{2}-M   ,\frac{1}{2}-M  ;1; \frac{ ( z-1)^2 -y^2 }{( z+1)^2 -y^2 }\Big) \Bigg]\Bigg|   
\end{eqnarray*}
that is bounded since  $F \Big(\frac{1}{2}   ,\frac{1}{2}   ;1; 0\Big)= F \Big(-\frac{1}{2}  ,\frac{1}{2}   ;1; 0\Big)=1$ and, consequently, from (\ref{2.2new}) by (\ref{1.10}) we derive
\begin{eqnarray*} 
&  & 
 \int_{ 0}^{1} \phi (t)^{a} s^{a}
\big|  K_0(\phi (t)s,t;M)\big|   \phi (t)\,  ds \\
& \leq &
%4 ^ {-\Re M}  z^{ - \Re M}  \int_{ 0}^{z-1 }  y^{a}
%  \big((1+z)^2 - y^2\big)^{  \Re M    } \frac{1}{ [( z-1)^2 -  y^2]\sqrt{(z+ 1)^2 - y^2} }\\
%&   &
%\times  \Bigg|\Bigg[  \big(  z -1 +M(z^{ 2 } -      1 -  y^2) \big) 
%F \Big(\frac{1}{2}-M   ,\frac{1}{2}-M  ;1; \frac{ ( z-1)^2 -y^2 }{( z+1)^2 -y^2 }\Big) \\
%&  &
%\hspace{1cm}  +   \big( 1-z^{2  }+  y^2 \big)\Big( \frac{1}{2}+M\Big)
%F \Big(-\frac{1}{2}-M   ,\frac{1}{2}-M  ;1; \frac{ ( z-1)^2 -y^2 }{( z+1)^2 -y^2 }\Big) \Bigg]\Bigg|    \,  d  y \\
%& \leq &
%4 ^ {-\Re M}  z^{ - \Re M}  \int_{ 0}^{z-1 }  y^{a}
%  \big((1+z)^2 - y^2\big)^{  \Re M    } \frac{1}{  \sqrt{(z+ 1)^2 - y^2} }  \,  d  y \\
% & \leq &
4 ^ {-\Re M}  z^{ - \Re M}  \int_{ 0}^{z-1 }  y^{a}
  \big((1+z)^2 - y^2\big)^{  \Re M -\frac{1}{2}   }   \,  d  y \\
& \leq &
4 ^ {-\Re M}  z^{ - \Re M}  \frac{1}{1+a}(z-1)^{1+a} (z+1)^{2 \Re M-1} F\left(\frac{1+a}{2},\frac{1}{2}-\Re M;\frac{3+a}{2};\frac{(z-1)^2}{(z+1)^2}\right) \\
& \lesssim &
   z^{ - \Re M}   (z-1)^{1+a} (z+1)^{2 \Re M-1}  , \qquad 1 <  z \leq N \,.  
\end{eqnarray*} 
Finally
\[ 
 \int_{ 0}^{1} \phi (t)^{a} s^{a}
\big|  K_0(\phi (t)s,t;M)\big|   \phi (t)\,  ds \\ 
  \lesssim  
   z^{ - \Re M}   (z-1)^{1+a} (z+1)^{2 \Re M-1}  , \qquad 1 <  z \leq N \,.  
\] 
Thus, we can restrict ourselves to the  case of large $   z \geq N$. 
We fix $\varepsilon \in (0,1)$ and   divide the domain of integration into  two zones,  
\begin{eqnarray*}
Z_1(\varepsilon, z) 
& := &
\left\{ (z,y) \,\Big|\, \frac{ (z-1)^2 -y^2   }{ (z+1)^2 -y^2 } \leq \varepsilon,\,\, 0 \leq y \leq z-1 \right\} ,\\
Z_2(\varepsilon, z) 
& := &
\left\{ (z,y) \,\Big|\, \varepsilon \leq  \frac{ (z-1)^2 -y^2   }{ (z+1)^2 -y^2 },\,\, 0 \leq y \leq z-1  \right\}\,.
\end{eqnarray*}
Furthermore, we  split the integral into  two parts:
\begin{eqnarray*}
\int_{ 0}^{z-1} y^{a} 
\big|  K_0(y,t;M)\big|    \,  dy
& = &
\int_{ (z,r) \in Z_1(\varepsilon, z)   }  y^{a} 
\big|  K_0(y,t;M)\big|    \,  dy     
+ \int_{ (z,r) \in Z_2(\varepsilon, z)   }  y^{a} 
\big|  K_0(y,t;M)\big|    \,  dy \,.
\end{eqnarray*} 
 In the first zone we have (7.11)~\cite{Yag_Galst_CMP}:
\begin{eqnarray*}
F\Big(\frac{1}{2}-M,\frac{1}{2}-M;1; \varepsilon   \Big) 
 &  =  &
1+\left(\frac{1}{4}-M \right)^2 \varepsilon +O\left(\varepsilon ^2\right), \\
F\Big(-\frac{1}{2}-M,\frac{1}{2}-M;1; \varepsilon    \Big) 
 & = &
1+\left(M^2-\frac{1}{4}\right)\varepsilon  +O\left(\varepsilon ^2\right) . 
\end{eqnarray*}
We use the last formulas to estimate the terms of (\ref{2.2new}) containing the hypergeometric functions:
\begin{eqnarray*} 
\hspace{-0.2cm} &  & 
\Bigg|\Bigg[  \big(  z -1 +M(z^{ 2 } -      1 -  y^2) \big) 
F \Big(\frac{1}{2}-M   ,\frac{1}{2}-M  ;1; \frac{ ( z-1)^2 -y^2 }{( z+1)^2 -y^2 }\Big) \\
\hspace{-0.2cm}&  &
\hspace{0.5cm}  +   \big( 1-z^{2  }+  y^2 \big)\Big( \frac{1}{2}+M\Big)
F \Big(-\frac{1}{2}-M   ,\frac{1}{2}-M  ;1; \frac{ ( z-1)^2 -y^2 }{( z+1)^2 -y^2 }\Big) \Bigg]\Bigg|    \\
\hspace{-0.2cm}& = &
\Bigg|\Bigg[  \big(  z -1 +M(z^{ 2 } -      1 -  y^2) \big) 
\Big[ 1+\big(M^2-M+\frac{1}{4}\big) \frac{ ( z-1)^2 -y^2 }{( z+1)^2 -y^2 }+O\Big(\Big(\frac{ ( z-1)^2 -y^2 }{( z+1)^2 -y^2 }\Big)^2\Big)\Big] \\
\hspace{-0.2cm}&  &
 +   \big( 1-z^{2  }+  y^2 \big)\Big( \frac{1}{2}+M\Big)\Big(1+\big(M^2-\frac{1}{4}\big) \frac{ ( z-1)^2 -y^2 }{( z+1)^2 -y^2 }
+O\Big(\Big(\frac{ ( z-1)^2 -y^2 }{( z+1)^2 -y^2 }\Big)^2\Big)\Big) \Bigg]\Bigg|   \\
\hspace{-0.2cm}& \leq  &
%\Bigg|  \frac{1}{2}\left(y^2-z^2+2 z-1\right) \\
%&  &
%+\frac{1}{8}   \left(12 M^2 y^2-12 M^2 z^2+8 M^2 z+4 M^2-4 M y^2+4 M z^2-8 M z+4 M-y^2+z^2+2 z-3\right)\\
%&  &
%\times \frac{ ( z-1)^2 -y^2 }{( z+1)^2 -y^2 } 
%+ \left(     z -1 +M(z^{ 2 } -      1 -  y^2)      \right) O\left(\left(\frac{ ( z-1)^2 -y^2 }{( z+1)^2 -y^2 }\right)^2\right)\Bigg|\\
%&  &
%+  \big( 1-z^{2  }+  y^2 \big)\Big( \frac{1}{2}+M\Big)     O\left(\left(\frac{ ( z-1)^2 -y^2 }{( z+1)^2 -y^2 }\right)^2\right)\Bigg|  \\
%& = &
\frac{1}{2} \left((z-1)^2-y^2\right)\\
\hspace{-0.2cm}&  &
+\frac{1}{8}  \Bigg|  (2 M-1) \left(6 M (y^2- z^2)+4 M z+2 M+y^2-z^2-2 z+3\right)\Bigg|  \frac{ ( z-1)^2 -y^2 }{( z+1)^2 -y^2 }\\
\hspace{-0.2cm}&  &
+\Bigg|\left(     z -1 +M(z^{ 2 } -      1 -  y^2)      \right) O\Big(\Big(\frac{ ( z-1)^2 -y^2 }{( z+1)^2 -y^2 }\Big)^2\Big)\\
\hspace{-0.2cm}&  &
+  \big( 1-z^{2  }+  y^2 \big)\Big( \frac{1}{2}+M\Big)     O\Big(\Big(\frac{ ( z-1)^2 -y^2 }{( z+1)^2 -y^2 }\Big)^2\Big)\Bigg|  \,.
\end{eqnarray*}
Hence, we have to consider the following three integrals, which can be easily evaluated and estimated (see, also, (\ref{1.10}))
\begin{eqnarray*} 
A_1
& := &
\int_{ 0}^{z-1 }  y^{a}
  \big((1+z)^2 - y^2\big)^{  \Re M  } \frac{1}{ [( z-1)^2 -  y^2]\sqrt{(z+ 1)^2 - y^2} }  \left((z-1)^2-y^2\right)     \,  d  y  \\
%& = &
%\int_{ 0}^{z-1 }  y^{a}
%  \big((1+z)^2 - y^2\big)^{  \Re M    -\frac{1}{2}  }    \,  d  y \\
%& = &
% \frac{1}{a+1}(z-1)^{a+1} (z+1)^{2 \Re M-1} \,  F \left(\frac{a+1}{2},\frac{1}{2}-\Re M;\frac{a+3}{2};\frac{(z-1)^2}{(z+1)^2}\right)\\
& \lesssim &
 \frac{1}{a+1}(z-1)^{a+1} (z+1)^{2 \Re M-1} \,, \\ 
A_2
& := &
\int_{ 0}^{z-1 }  y^{a}
  \big((1+z)^2 - y^2\big)^{  \Re M    } \frac{1}{ [( z-1)^2 -  y^2]\sqrt{(z+ 1)^2 - y^2} }\\
&   &
\times  \Bigg|    (2 M-1) \left(6 M (y^2- z^2)+4 M z+2 M+y^2-z^2-2 z+3\right)\Bigg|  \frac{ ( z-1)^2 -y^2 }{( z+1)^2 -y^2 }     \,  d  y \,.
\end{eqnarray*}
Consider the real and imaginary parts of the function $Z=Z(y) $:
\begin{eqnarray*}  
\Re Z(y)& = &
 \frac{\left(6 \Re M (y^2- z^2)+4 \Re M z+2 \Re M+y^2-(z+1)^2 +4\right)}{( z+1)^2 -y^2}\,,\\
\Im Z(y)& = & 
\frac{\left(6 \Im M (y^2- z^2)+4 \Im M z+2 \Im M\right)}{( z+1)^2 -y^2} \,.
\end{eqnarray*}
The function $Z=Z(y) $  takes the following values at two end points
\[
Z(z-1)
  =  
 -\frac{(2 M+1) (z-1)}{z},\quad Z(0)=-\frac{(z-1) (M (6 z+2)+z+3)}{(z+1)^2}\,.
\]
Then with $N:= \Im M$, $ \widetilde{M}:= \Re M  $  and $b=y^2$ we have
\begin{eqnarray*}  
&  &
\frac{d}{db}\left[(\Re Z(y))^2+ (\Im Z(y))^2 \right]\\
& = &
\frac{8}{(( z+1)^2 -y^2)^3}\Bigg[\left(12 b \widetilde{M}^2+8 b \widetilde{M}+12 b N^2+b+4 \widetilde{M}^2+8 \widetilde{M}+4 N^2+3\right)\\
&  &
+z \left(24 b \widetilde{M}^2+4 b \widetilde{M}+24 b N^2+16 \widetilde{M}^2+12 \widetilde{M}+16 N^2-2\right)\\
&  &
+z^2 \left(4 \widetilde{M}^2-16 \widetilde{M}+4 N^2-1\right)+z^3 \left(-24 \widetilde{M}^2-4 \widetilde{M}-24 N^2\right)\Bigg]\\
&  < &
 0 \quad \mbox{\rm for large} \quad z \quad \mbox{\rm and all} \quad y \in[0,z-1].
\end{eqnarray*} 
Indeed,
\begin{eqnarray*} 
\frac{d}{db}\left[(\Re Z(y))^2+ (\Im Z(y))^2 \right]\Bigg|_{b=0} 
& = &
-(z+1)^{-6}\Big[  8 (z-1)  ( z^2  (24 \widetilde{M}^2+4 \widetilde{M}+24 N^2 )+z  (20 \widetilde{M}^2\\
&  &
+20 \widetilde{M}+20 N^2+1 )+ (4 \widetilde{M}^2+8 \widetilde{M}+4 N^2+3 ) )\Big]
\end{eqnarray*}
%\begin{eqnarray*}  
%&  &
%\frac{d}{db}\left[(\Re Z(y))^2+ (\Im Z(y))^2 \right]\Bigg|_{b=0}\\
%& = &
%-\frac{8 (z-1) \left( z^2 \left(24 M^2+4 M+24 N^2\right)+z \left(20 M^2+20 M+20 N^2+1\right)+\left(4 M^2+8 M+4 N^2+3\right)\right)}{(z+1)^6}
%\end{eqnarray*}
and
\begin{eqnarray*}  
&  &
\frac{d}{db}\left[(\Re Z(y))^2+ (\Im Z(y))^2 \right]\Bigg|_{b=(z-1)^2}\\
& = &
-2^{-1} z^{-3}\Big[  (z-1)  (\widetilde{M}^2 (8 z+4)+4 \widetilde{M} (z+1)+N^2 (8 z+4)+1 )\Big]\,.
\end{eqnarray*} 
%\begin{eqnarray*}  
%&  &
%\frac{d}{db}\left[(\Re Z(y))^2+ (\Im Z(y))^2 \right]\Bigg|_{b=(z-1)^2}\\
%& = &
%-\frac{(z-1) \left(M^2 (8 z+4)+4 M (z+1)+N^2 (8 z+4)+1\right)}{2 z^3}\,.
%\end{eqnarray*} 
Hence,
\begin{eqnarray*}  
A_2
& \lesssim &
 (z-1)^{a+1} (z+1)^{2 \Re M-1} \,.
\end{eqnarray*}
In the first zone we have
\begin{eqnarray*} 
A_3
& := &  
\int_{ 0}^{z-1 }  y^{a}
  \big((1+z)^2 - y^2\big)^{  \Re M    } \frac{\left|     z -1 +M(z^{ 2 } -      1 -  y^2)      \right|}{ [( z-1)^2 -  y^2]\sqrt{(z+ 1)^2 - y^2} }  \Bigg|   \frac{ ( z-1)^2 -y^2 }{( z+1)^2 -y^2 } \Bigg|^2    \,  d  y \\
&   &  
+ \int_{ 0}^{z-1 }  y^{a}
  \big((1+z)^2 - y^2\big)^{  \Re M    } \frac{\big(z^{2  }- 1-   y^2 \big)\Big| \frac{1}{2}+M\Big|}{ [( z-1)^2 -  y^2]\sqrt{(z+ 1)^2 - y^2} }  \Bigg|   \frac{ ( z-1)^2 -y^2 }{( z+1)^2 -y^2 } \Bigg|^2    \,  d  y\,. 
\end{eqnarray*}
Then 
\begin{eqnarray*} 
A_3
& \lesssim  &  
\int_{ 0}^{z-1 }  y^{a}
  \big((1+z)^2 - y^2\big)^{  \Re M    } \frac{ |     z -1| +|z^{ 2 } -      1 -  y^2 |}{ \sqrt{(z+ 1)^2 - y^2} }      \frac{ [( z-1)^2 -y^2] }{(( z+1)^2 -y^2)^2 }      \,  d  y \\
& \lesssim  &  
%z^2 \int_{ 0}^{z-1 }  y^{a}
%  \big((1+z)^2 - y^2\big)^{  \Re M    } \frac{ 1}{ \sqrt{(z+ 1)^2 - y^2} }      \frac{ [( z-1)^2 -y^2] }{(( z+1)^2 -y^2)^2 }      \,  d  y \\
%& \lesssim  &  
z^2 \int_{ 0}^{z-1 }  y^{a}
  \big((1+z)^2 - y^2\big)^{  \Re M -\frac{3}{2}   }          \,  d  y \\
& = &  
z^2 \frac{1}{a+1}(z-1)^{a+1} (z+1)^{2 \Re M-3} \, F\left(\frac{a+1}{2},\frac{3}{2}-M;\frac{a+3}{2};\frac{(z-1)^2}{(z+1)^2}\right)\\ 
& \lesssim &
 (z-1)^{a+1} (z+1)^{  2\Re M-1} \times 
\cases{ 1 \quad \mbox{\rm if } \quad   \Re \, M>1/2 \cr
\ln z  \quad \mbox{\rm if } \quad \Re \,M=1/2 \cr
z^{\frac{1}{2}-\Re M }   \quad \mbox{\rm if } \quad  1/2>\Re \,M  } \,.
\end{eqnarray*}
Finally in the first zone we obtain
\begin{eqnarray*}
 \int_{ (z,r) \in Z_1(\varepsilon, z)   }  y^{a} 
\big|  K_0(y,t;M)\big|    \,  dy   
& \lesssim &
 (z-1)^{a+1} (z+1)^{  \Re M-1} \times 
\cases{ 1 \quad \mbox{\rm if } \quad   \Re \, M>1/2 \cr
\ln z  \quad \mbox{\rm if } \quad \Re \,M=1/2 \cr
z^{\frac{1}{2}-\Re M }   \quad \mbox{\rm if } \quad  1/2>\Re \,M  } \,.
\end{eqnarray*}
In the second zone   
\[
\varepsilon \leq  \frac{ (z-1)^2 -y^2   }{ (z+1)^2 -y^2 } \leq 1   \quad \mbox{\rm implies}  \quad 
\frac{ 1  }{ (z-1)^2 -y^2 }  \leq  \frac{ 1   }{ \varepsilon[(z+1)^2 -y^2] }\,\,.
\]
Further, the hypergeometric functions for $\Re M>0$ obey the estimates
\[
\left| F\Big(-\frac{1}{2}-M,\frac{1}{2}-M;1; \zeta      \Big) \right|  \leq C \,\,  \mbox{\rm and}  \,\,   
\left| F\Big(\frac{1}{2}-M,\frac{1}{2}-M;1; \zeta   \Big) \right|  \leq C_M   \,\, \, \mbox{\rm for all}\,\,   \zeta  \in [0 ,1) .
\]
This allows us to  estimate  the integral over the second zone:  
\begin{eqnarray*} 
&  &
\int_{ 0,(z,r) \in Z_2(\varepsilon, z) }^{z-1 }  y^{a}
  \big((1+z)^2 - y^2\big)^{  \Re M    } \frac{1}{ [( z-1)^2 -  y^2]\sqrt{(z+ 1)^2 - y^2} }\\
&   &
\times  \Bigg|\Bigg[  \big(  z -1 +M(z^{ 2 } -      1 -  y^2) \big) 
F \Big(\frac{1}{2}-M   ,\frac{1}{2}-M  ;1; \frac{ ( z-1)^2 -y^2 }{( z+1)^2 -y^2 }\Big) \\
&  &
\hspace{1cm}  +   \big( 1-z^{2  }+  y^2 \big)\Big( \frac{1}{2}+M\Big)
F \Big(-\frac{1}{2}-M   ,\frac{1}{2}-M  ;1; \frac{ ( z-1)^2 -y^2 }{( z+1)^2 -y^2 }\Big) \Bigg]\Bigg|    \,  d  y  \\
& \lesssim &
%\int_{ 0,(z,r) \in Z_2(\varepsilon, z) }^{z-1 }  y^{a}
%  \big((1+z)^2 - y^2\big)^{  \Re M  -\frac{3}{2}  }  \\
%&   &
%\times  \Bigg|\Bigg[  \big(  z -1 +M(z^{ 2 } -      1 -  y^2) \big) 
%F \Big(\frac{1}{2}-M   ,\frac{1}{2}-M  ;1; \frac{ ( z-1)^2 -y^2 }{( z+1)^2 -y^2 }\Big) \\
%&  &
%\hspace{1cm}  +   \big( 1-z^{2  }+  y^2 \big)\Big( \frac{1}{2}+M\Big)
%F \Big(-\frac{1}{2}-M   ,\frac{1}{2}-M  ;1; \frac{ ( z-1)^2 -y^2 }{( z+1)^2 -y^2 }\Big) \Bigg]\Bigg|    \,  d  y \\
%& \lesssim &
z^{ 2 } \int_{ 0}^{z-1 }  y^{a}
  \big((1+z)^2 - y^2\big)^{  \Re M  -\frac{3}{2}  }     \,  d  y \\
& = &
(a+1)^{-1}z^{ 2 }   (z-1)^{a+1} (z+1)^{2 \Re M-3}  F \left(\frac{a+1}{2},\frac{3}{2}-\Re M;\frac{a+3}{2};\frac{(z-1)^2}{(z+1)^2}\right)\,.
\end{eqnarray*}
According to  Lemma~A1~\cite{Y2017} with $a>-1$ we have
\begin{eqnarray*}
%\label{F32a}
&  &
\lim_{z \to \infty}  \,  F \left(\frac{a+1}{2},\frac{3}{2}-M;\frac{a+3}{2};\frac{(z-1)^2}{(z+1)^2}\right)
= \frac{\Gamma (\frac{a+3}{2})\Gamma (M - \frac{ 1}{2} )}{\Gamma (\frac{a }{2} +M)}\quad \mbox{\rm if } \quad    M>1/2  \,,\\
&  &
\lim_{z \to \infty}  \,\frac{1}{\ln z}  F \left(\frac{a+1}{2},\frac{3}{2}-M;\frac{a+3}{2};\frac{(z-1)^2}{(z+1)^2}\right)
= \frac{1 + a}{2} \quad \mbox{\rm if } \quad M=1/2 \,, \\
&  &
%\label{F32}
\lim_{z \to \infty} z^{M-\frac{1}{2}} \,  F \left(\frac{a+1}{2},\frac{3}{2}-M;\frac{a+3}{2};\frac{(z-1)^2}{(z+1)^2}\right)
=  2^{2 M-1}\frac{1+a }{1-2 M }   \quad \mbox{\rm if } \quad  1/2> M  \,.
\end{eqnarray*}  
Hence,
\begin{eqnarray*}
&  &
 \int_{ (z,r) \in Z_2(\varepsilon, z)   }  y^{a} 
\big|  K_0(y,t;M)\big|    \,  dy    \\ 
& \lesssim &
%z^{2-\Re M }(z-1)^{a+1} (z+1)^{ 2 \Re M-3} \times 
%\cases{ 1 \quad \mbox{\rm if } \quad   \Re \, M>1/2 \cr
%\ln z  \quad \mbox{\rm if } \quad \Re \,M=1/2 \cr
%z^{\frac{1}{2}-\Re M }   \quad \mbox{\rm if } \quad  1/2>\Re \,M  } \\ 
%& \lesssim &
 (z-1)^{a+1} (z+1)^{  \Re M-1} \times 
\cases{ 1 \quad \mbox{\rm if } \quad   \Re \, M>1/2 \cr
\ln z  \quad \mbox{\rm if } \quad \Re \,M=1/2 \cr
z^{\frac{1}{2}-\Re M }   \quad \mbox{\rm if } \quad  1/2>\Re \,M  } \,.
\end{eqnarray*} 
Then we combine the estimates obtained in two zones
\begin{eqnarray*}
\int_{ 0}^{z-1} y^{a} 
\big|  K_0(y,t;M)\big|    \,  dy 
& = &
\int_{ (z,r) \in Z_1(\varepsilon, z)   }  y^{a} 
\big|  K_0(y,t;M)\big|    \,  dy     
+ \int_{ (z,r) \in Z_2(\varepsilon, z)   }  y^{a} 
\big|  K_0(y,t;M)\big|    \,  dy \\
& \lesssim &
(z-1)^{a+1} (z+1)^{  \Re M-1}    
\cases{ 1 \quad \mbox{\rm if } \quad   \Re \, M>1/2 \,,\cr
\ln z  \quad \mbox{\rm if } \quad \Re \,M=1/2 \,,\cr
z^{\frac{1}{2}-\Re M }   \quad \mbox{\rm if } \quad  1/2>\Re \,M   \,.} 
\end{eqnarray*}
The lemma is proved. \hfill $\square$

\section{%$\bf  B^{s,q}_{p}- B^{s',q}_{p'}$ 
Estimates for the equation without a source}
\label{S30}

\setcounter{equation}{0}
\renewcommand{\theequation}{\thesection.\arabic{equation}}

Let $\varphi _j= \varphi (2^{-j} \xi )  $, $j>0$, and $\varphi _0=1-\sum_{j=1}^\infty\varphi_j  $, where $\varphi \in C_0^\infty ({\mathbb R}^n) $ 
with $\varphi \geq 0 $  and \,\,supp$\,\varphi \subseteq \{ \xi \in {\mathbb R}^n\,;\, 1/2 <|\xi|< 2\}$, is that 
\[
\sum_{-\infty}^\infty \varphi  (2^{-j} \xi ) =1, \quad \xi \not=0\,.
\]
The norm $\|v\|_{B^{s,q}_{p}} $ of the Besov space $ B^{s,q}_{p}$ is defined as follows
\begin{eqnarray*}
&   &
\|v\|_{B^{s,q}_{p}}= \left(\sum_{j=0}^\infty \left( 2^{js} \| {\mathcal F}^{-1}\left( \varphi _j\hat{v}  \right) \|_{p} \right)^q \right) ^{1/q}\,,
\end{eqnarray*} 
where $\hat{v}  $ is the Fourier transform of $v$. The following theorem by Brenner~\cite{Brenner} is crucial   for   this and the next sections.

\begin{theorem} \mbox{\rm (Brenner~\cite{Brenner})}
\label{TBrenner}
Let $A=A(x,D)$ be a second order negative elliptic differential operator with real $C^\infty$-coefficients such that 
$ A(x,D) =A(\infty,D)$ for $|x|$ large enough. Let $u(t) = G_0(t)g_0+ G_1(t)g_1$ be the solution of 
\begin{eqnarray}
\label{2.1}
&   &
\partial_t^2 u - A(x,D)u =0, \quad x \in {\mathbb R}^n, \quad t \geq 0, \\
\label{2.2}
&  &
u (x,0) = g_0 (x),\quad   u_t (x,0)= g_1(x),\quad x \in {\mathbb R}^n\,.
\end{eqnarray}
Then for each $T<\infty$ there is a constant\, $C=C(T)$ such that if \, $(n+1)\delta \leq \nu +s-s' $,\, $\nu =0,1$, 
\begin{eqnarray}
\label{Blplq}
&   &
\| G_\nu (t) g\| _{B^{s',q}_{p'}} \leq C(T)t^{\nu +s-s'-2n\delta } \|g\|_{B^{s,q}_{p}}, \quad 0 < t \leq T  \,.
\end{eqnarray}
Here $s,s' \geq 0$, $q\geq 1 $, $1\leq p\leq 2$, $1/p+1/p'=1$, and $\delta =1/p-1/2 $. 
\end{theorem}
The estimate (\ref{Blplq}) can be also written as
\[
\| G_\nu (t) g\| _{B^{s',q}_{p'}} \leq C(T)t^{\nu +s-s'-n\left(1/p-1/p' \right) } \|g\|_{B^{s,q}_{p}}, \quad 0 < t \leq T, \quad \nu =0,1 \,.
\]
\begin{remark}
In the case of $A(x,D)=\Delta  $ the constant $C(T)$ can be chosen independent of $T$, that is, % due to the scaling arguments.  
\[
\| G_\nu (t) g\| _{B^{s',q}_{p'}} \leq   c t^{\nu +s-s'-n\left(1/p-1/p' \right) } \|g\|_{B^{s,q}_{p}} \quad \mbox{for all} \quad t \in (0,\infty), \quad \nu =0,1 \,.
\]
\end{remark}
(See, also, e.g. \mbox{\rm \cite{Brenner,Pecher,Yagdjian_CPDE_2006}}.)  
\begin{theorem} 
\label{T13.2}
Suppose that for the equation (\ref{2.1}) the conditions of Theorem~\ref{TBrenner} are fulfilled.  Assume that $s,s' \geq 0$, $q\geq 1 $, $1\leq p\leq 2$, $1/p+1/p'=1$, and $\delta =1/p-1/2 $, $(n+1)\delta \leq s-s' $, $-1 <s-s'-2n\delta  $.
Denote $a:=s-s'-2n\delta$. 

Then for each $T<\infty$ there is a constant\, $C=C(T)$ such that the solution  $\Phi =\Phi (x,t)$ of the Cauchy problem 
\begin{equation}
\label{1.7}
  \Phi _{tt} -  n   \Phi _t - e^{2 t} A(x,D) \Phi  + m^2\Phi =  0\,, \quad \Phi  (x,0)= \varphi_0 (x)\, , \quad \Phi  _t(x,0)=\varphi_1 (x)\,,
\end{equation}
where $ \Re M>0$, satisfies the following  estimate
\begin{eqnarray*}  
\| \Phi  (x,t) \| _{B^{s',q}_{p'}}  
&   \leq   &  
C(T)\|\varphi _0 \|_{B^{s,q}_{p}} e^{\frac{n}{2}t}(e^{t}-1)^{a}\Bigg(   e^{-\frac{1}{2}t}  
+  
(e^{t }-1) (e^t+1)^{  \Re M-1}\Bigg[ e^{-\Re Mt } (e^{t }+1)^{  \Re M }\\
&  &
+      
\cases{ 1 \quad \mbox{\rm if } \quad   \Re \, M>1/2 \cr  
t^{\sgn|\Im M|}+ e^{(\frac{1}{2}-\Re M)t  }   \quad \mbox{\rm if } \quad \Re \,M  \leq  1/2 } \Bigg] \Bigg)\nonumber \\ 
&  & 
\,+ C(T)\|\varphi _1 \|_{B^{s,q}_{p}}e^{\frac{n}{2}t}   e^{-\Re Mt }(e^{t }-1)^{a+1} (e^{t }+1)^{2 \Re M-1}
  \, \,\, \mbox{  for all}\,\,\, 0<t< T.
\end{eqnarray*}
If $A(x,D)=\Delta  $, then the constant $C(T)$ is   independent of $T$, that is, the estimate holds with $T=\infty$. 
\end{theorem} 
\begin{corollary}
\label{C1.4}
The solution  $\Phi =\Phi (x,t)$ of the Cauchy problem  
(\ref{1.7}) satisfies the following  estimates
\begin{eqnarray*}  
\| \Phi  (x,t) \| _{B^{s',q}_{p'}}  
&   \leq   &  
C(T) \|\varphi _0 \|_{B^{s,q}_{p}} e^{(\frac{n}{2}+a+\Re M)t}    
\cases{ 1 \quad \mbox{\rm if } \quad   \Re \, M>1/2 \cr  
t^{\sgn|\Im M|}+ e^{(\frac{1}{2}-\Re M)t  }   \quad \mbox{\rm if } \quad \Re \,M  \leq  1/2 } \nonumber \\ 
&  & 
\,+ C(T)\|\varphi _1 \|_{B^{s,q}_{p}}e^{(\frac{n}{2}+a+\Re M)t}
, \, \,\, \mbox{  for all}\,\,\,  t \in (1,T), \\
%\label{phi10n2}  
\hspace{-1.2cm}
\| \Phi  (x,t) \| _{B^{s',q}_{p'}} 
&  \leq   &        
C(T)t^a \|\varphi _0 \|_{B^{s,q}_{p}}        
+   t^{a+1} \|\varphi _1 \|_{B^{s,q}_{p}} \, \,\, \mbox{  for all}\,\,  t \in (0,1)\,.
\end{eqnarray*} 
If $A(x,D)=\Delta  $, then the constant $C(T)$ is independent of $T$, that is, the estimates hold with $T=\infty$. 
\end{corollary}
\medskip

\noindent
{\bf Proof of Theorem~\ref{T13.2}.}  The proofs for the cases of the Laplace operator $\Delta  $ and  the operator $A(x,\partial_x)  $
are identical and we set now $A(x,\partial_x)=\Delta  $.  First 
we consider the case of $\varphi  _1=0 $. Then according to  (\ref{Phy})
\begin{eqnarray*} 
&  &
\Phi  (x,t) \\
 & =  &
 e^{\frac{n-1}{2}t} v_{\varphi_0}  (x, \phi (t))
+ \,  e^{\frac{n}{2}t}\int_{ 0}^{1} v_{\varphi_0}  (x, \phi (t)s)\big(2  K_0(\phi (t)s,t;M)- nK_1(\phi (t)s,t;M)\big)\phi (t)\,  ds  \nonumber 
\end{eqnarray*}
and, consequently, 
\begin{eqnarray} 
\label{2.9}
\| \Phi  (x,t) \| _{B^{s',q}_{p'}}  
&   \leq &  
e^{\frac{n-1}{2}t}\|  v_{\varphi_0}  (x, \phi (t)) \| _{B^{s',q}_{p'}}   \\
&  &
+ \,  e^{\frac{n}{2}t}\int_{ 0}^{1} \| v_{\varphi_0}  (x, \phi (t)s)\| _{B^{s',q}_{p'}}  \big|2  K_0(\phi (t)s,t;M)- nK_1(\phi (t)s,t;M)\big|\phi (t)\,  ds\,. \nonumber
\end{eqnarray}
If $n \geq 2$, then according to Theorem~\ref{TBrenner}, for the solution $v = v (x,t)$ of the Cauchy problem (\ref{2.1})-(\ref{2.2}) 
with $\varphi (x) \in C_0^\infty({\mathbb R}^n)$ one has  the estimate (\ref{Blplq}) 
provided that \,  $s,s' \geq 0$, $q\geq 1 $, $1\leq p\leq 2$, $1/p+1/p'=1$, and $\delta =1/p-1/2 $, $(n+1)\delta \leq  s-s' $.
Hence,  
\[ 
 \|  v_{\varphi_0}  (x, \phi (t)) \| _{B^{s',q}_{p'}}
 \leq  
C\phi (t)^{s-s'-2n\delta}\|\varphi _0 \|_{B^{s,q}_{p}}  \quad \mbox{\rm for all} \quad  t >0,
\]
where $\phi (t)= e^{t}-1$. Consequently, for the first term of the right-hand side of (\ref{2.9}) we have
\[  
e^{\frac{n-1}{2}t} \| v_{\varphi_0}  (x, \phi (t)) \| _{B^{s',q}_{p'}}
  \leq   
C  e^{\frac{n-1}{2}t}(e^{t}-1)^{s-s'-2n\delta}\|\varphi _0 \|_{B^{s,q}_{p}}  \quad \mbox{\rm for all} \quad  t >0\,.
\]
For the second  term of (\ref{2.9}) we obtain
\begin{eqnarray*} 
&  &
e^{\frac{n}{2}t}\int_{ 0}^{1} \|  v_{\varphi_0}  (x, \phi (t)s) \| _{B^{s',q}_{p'}} \big|2  K_0(\phi (t)s,t;M)- nK_1(\phi (t)s,t;M)\big|\phi (t)\,  ds\\
& \leq &
 \|\varphi _0 \|_{B^{s,q}_{p}}e^{\frac{n}{2}t} \int_{ 0}^{1} \phi (t)^{s-s'-2n\delta} s^{s-s'-2n\delta} \left( \big|2  K_0(\phi (t)s,t;M)\big|+ n\big|K_1(\phi (t)s,t;M)\big|\right) \phi (t)\,  ds \,.
\end{eqnarray*}
Denote $a:=s-s'-2n\delta$. We have to estimate the following two terms of the last inequality:
\begin{eqnarray*} 
&  & 
 \int_{ 0}^{1} \phi (t)^{a} s^{a}
\big|  K_i(\phi (t)s,t;M)\big|   \phi (t)\,  ds, \quad i=0,1 \,,
\end{eqnarray*}
where  $\phi (t)= e^{t}-1$ and $t>0$. To complete the estimate of the second term  of (\ref{2.9}) we   apply  Lemma~\ref{L2.3} and Lemma~\ref{L1.2}. 
 Thus, if $\varphi _1=0 $, then from  (\ref{2.9})   we derive
\begin{eqnarray*}   
\| \Phi  (x,t) \| _{B^{s',q}_{p'}}  
&   \lesssim  &  
   e^{\frac{n-1}{2}t}(e^{t}-1)^{a}\|\varphi _0 \|_{B^{s,q}_{p}}   \\
&  &
+ \|\varphi _0 \|_{B^{s,q}_{p}}e^{\frac{n}{2}t} 
\Bigg\{ \int_{ 0}^{1} \phi (t)^{a} s^{a} \left( \big|   K_0(\phi (t)s,t;M)\big|+  \big|K_1(\phi (t)s,t;M)\big|\right) \phi (t)\,  ds \Bigg\}\\ 
&   \lesssim  &  
\|\varphi _0 \|_{B^{s,q}_{p}} \Bigg(   e^{\frac{n-1}{2}t}(e^{t}-1)^{a}  
+ e^{\frac{n}{2}t} 
\Bigg[ e^{-\Re Mt }(e^{t }-1)^{a+1} (e^{t }+1)^{2 \Re M-1}\\
&  &
+   (e^t-1)^{a+1} (e^t+1)^{  \Re M-1}   
\cases{ 1 \quad \mbox{\rm if } \quad   \Re \, M>1/2 \cr  
t^{\sgn|\Im M|}+ e^{(\frac{1}{2}-\Re M)t  }   \quad \mbox{\rm if } \quad \Re \,M  \leq  1/2 } \Bigg] \Bigg) \\
&   \lesssim  &  
\|\varphi _0 \|_{B^{s,q}_{p}} \Bigg(   e^{\frac{n-1}{2}t}(e^{t}-1)^{a}  
+ e^{\frac{n}{2}t} 
(e^{t }-1)^{a+1}\Bigg[ e^{-\Re Mt } (e^{t }+1)^{2 \Re M-1}\\
&  &
+    (e^t+1)^{  \Re M-1}   
\cases{ 1 \quad \mbox{\rm if } \quad   \Re \, M>1/2 \cr  
t^{\sgn|\Im M|}+ e^{(\frac{1}{2}-\Re M)t  }   \quad \mbox{\rm if } \quad \Re \,M  \leq  1/2 } \Bigg] \Bigg).
\end{eqnarray*}
% In particular,  from the last estimate   we obtain 
%\begin{eqnarray*}  
%\| \Phi  (x,t) \| _{B^{s',q}_{p'}}  
%&   \lesssim  &  
%\|\varphi _0 \|_{B^{s,q}_{p}} e^{(\frac{n}{2}+a+\Re M)t}    
%\cases{ 1 \quad \mbox{\rm if } \quad   \Re \, M>1/2 \cr  
%t^{\sgn|\Im M|}+ e^{(\frac{1}{2}-\Re M)t  }   \quad \mbox{\rm if } \,\, \Re \,M  \leq  1/2 } 
%, \, \,\, \forall \,  t \in (1,\infty),\nonumber \\
%\label{phi10n2}  
%\hspace{-1.2cm}
%\| \Phi  (x,t) \| _{B^{s',q}_{p'}} 
%&  \lesssim  &        
%t^a \|\varphi _0 \|_{B^{s,q}_{p}} \, \,\, \mbox{  for all}\,\,  t \in (0,1)\,.
%\end{eqnarray*} 
In the case of $\varphi _0=0 $  from (\ref{Phy}) we have 
\begin{eqnarray*} 
\| \Phi  (x,t) \| _{B^{s',q}_{p'}}
% & =  &
%2e^{\frac{n}{2}t}\| \int_{0}^1   v_{\varphi _1 } (x, \phi (t) s) 
%  K_1(\phi (t)s,t;M) \phi (t)\, ds \| _{B^{s',q}_{p'}}\\
 & \leq  &
2e^{\frac{n}{2}t} \int_{0}^1  \| v_{\varphi _1 } (x, \phi (t) s) \| _{B^{s',q}_{p'}}
 | K_1(\phi (t)s,t;M)| \phi (t)\, ds \\
 & \leq  &
2\|\varphi _1 \|_{B^{s,q}_{p}}e^{\frac{n}{2}t} \int_{0}^1  \phi (t)^{a}s^a
 | K_1(\phi (t)s,t;M)| \phi (t)\, ds 
, \quad  t>0\,.  
\end{eqnarray*}
Due to Lemma~\ref{L2.3}  we obtain the statement of the theorem. 
%\begin{eqnarray*}
%\| \Phi  (x,t) \| _{B^{s',q}_{p'}}
% & \lesssim  &
% e^{\frac{n}{2}t}   e^{-\Re Mt }(e^{t }-1)^{a+1} (e^{t }+1)^{2 \Re M-1}\|\varphi _1 \|_{B^{s,q}_{p}}
%, \quad \,\, t>0\, . 
%\end{eqnarray*}
% In particular, for large $t$ we obtain
%\begin{eqnarray*} 
%\| \Phi  (x,t) \| _{B^{s',q}_{p'}}
% & \lesssim  &
% e^{(\frac{n}{2}+a+ \Re M)t}   \|\varphi _1 \|_{B^{s,q}_{p}} \quad \mbox{\rm for all} \quad t \in [1,\infty)\,, \\ 
%\| \Phi  (x,t) \| _{B^{s',q}_{p'}}
% & \lesssim  &
% t^{a+1} \|\varphi _1 \|_{B^{s,q}_{p}} \quad \mbox{\rm for all} \quad t \in (0,1)\,.
%\end{eqnarray*} 
 Theorem is proved. \hfill $\square$
\begin{remark}
The estimates in the Lebesgue spaces for the case of large mass $m \geq n/2$ were discussed in \cite{Galstian_Trieste}.  
\end{remark}

\section{Estimates for the equation with a source}
\label{S2b}
\setcounter{equation}{0}
\renewcommand{\theequation}{\thesection.\arabic{equation}}

\begin{theorem} 
\label{T11.1}
Assume that for the equation (\ref{2.1}) the conditions of Theorem~\ref{TBrenner} are fulfilled.
Let $\Phi =\Phi (x,t)$ be a solution of the Cauchy problem  
\[ 
  \Phi _{tt} -  n   \Phi _t - e^{2 t} A(x,\partial_x) \Phi  + m^2\Phi =  f\,, \quad \Phi  (x,0)= 0\, , \quad \Phi  _t(x,0)=0\,.
\] 
Then for each $T<\infty$ there is a constant\, $C=C(T)$ such that the    solution $\Phi =\Phi (x,t)$ with $ \Re M >0 $    
satisfies the following  estimate:
\begin{eqnarray*} 
\| \Phi (x ,t)  \| _{B^{s',q}_{p'}}
 & \leq  & 
C(T)  e^{ t ( \frac{n}{2}+ \Re M +s-s'-n\left(\frac{1}{p} -\frac{1}{p'} \right))} 
\int_{ 0}^{t}   e^{-( \frac{n}{2}+\Re M )b}           \|f(x,b)\|_{B^{s,q}_{p}}  \,db   \,.   
\end{eqnarray*} 
for all $t \in(0,C(T))$, provided that $s,s' \geq 0$, $q\geq 1 $, $1\leq p\leq 2$, $1/p+1/p'=1$, and $\delta =1/p-1/2 $, $(n+1)\delta \leq s-s' $, $-1 <s-s'-2n\delta  $.
 If $A(x,D)=\Delta  $, then the constant $C(T)$ is independent of $T$, that is, the estimate holds with $T=\infty$.
\end{theorem}
\medskip

\noindent
{\bf Proof.}   From (\ref{Phy}) we have
\begin{eqnarray*}
\Phi  (x,t) 
 & =  &
2   e^{\frac{n}{2}t}\int_{ 0}^{t} db
  \int_{ 0}^{ e^{t}- e^{b}} dr  \,  e^{-\frac{n}{2}b} v(x,r ;b) E(r,t; 0,b;M)  \,.
\end{eqnarray*}
Denote $a:=s-s'-n\left(\frac{1}{p} -\frac{1}{p'} \right)>  -1 $, then according to (\ref{Blplq}) we can write
\begin{eqnarray*} 
&  & 
\| v(x,r ;b)   \| _{B^{s',q}_{p'}}  
\le C 
r^{a} \|f(x,b)\|_{B^{s,q}_{p}}  \quad \mbox{\rm for all} \quad  r >0\,.
\end{eqnarray*}
 Hence, 
\begin{eqnarray*} 
\| \Phi  (x,t) \| _{B^{s',q}_{p'}}  
 & \lesssim  &
   e^{\frac{n}{2}t}\int_{ 0}^{t} db
  \int_{ 0}^{ e^{t}- e^{b}} dr  \,  e^{-\frac{n}{2}b} \| v(x,r ;b) \| _{B^{s',q}_{p'}} | E(r,t; 0,b;M) |\\
%  & \lesssim  &
%    e^{\frac{n}{2}t}\int_{ 0}^{t} db
%   \int_{ 0}^{ e^{t}- e^{b}} dr  \,  e^{-\frac{n}{2}b} \| v(x,r ;b) \| _{B^{s',q}_{p'}} \Bigg|   e^{ -M(b+t) } \Big((e^{t }+e^{b})^2 - r^2\Big)^{-\frac{1}{2}+M    } \\
%  &  &
%  \times 
% F\Big(\frac{1}{2}-M   ,\frac{1}{2}-M  ;1; 
% \frac{ ( e^{t}-e^{b })^2 -r^2 }{( e^{t}+e^{b })^2 -r^2 } \Big)  \Bigg|\\
%  & \lesssim  &
%    e^{\frac{n}{2}t}\int_{ 0}^{t} db
%   \int_{ 0}^{ e^{t}- e^{b}} dr  \,  e^{-\frac{n}{2}b} r^{a} \|f(x,b)\|_{B^{s,q}_{p}}    e^{ -\Re M(b+t) } \Big((e^{t }+e^{b})^2 - r^2\Big)^{-\frac{1}{2}+\Re M    } \\
%  &  &
%  \times 
% \Bigg| F\Big(\frac{1}{2}-M   ,\frac{1}{2}-M  ;1; 
% \frac{ ( e^{t}-e^{b })^2 -r^2 }{( e^{t}+e^{b })^2 -r^2 } \Big)  \Bigg|\\
 & \lesssim  &
   e^{\frac{n}{2}t}\int_{ 0}^{t}  e^{ -\Re M(b+t) }  e^{-\frac{n}{2}b}\|f(x,b)\|_{B^{s,q}_{p}} \,db
  \int_{ 0}^{ e^{t}- e^{b}}  r^{a}       \\
 &  &
 \times 
\Big((e^{t }+e^{b})^2 - r^2\Big)^{-\frac{1}{2}+\Re M    }\Bigg| F\Big(\frac{1}{2}-M   ,\frac{1}{2}-M  ;1; 
\frac{ ( e^{t}-e^{b })^2 -r^2 }{( e^{t}+e^{b })^2 -r^2 } \Big)  \Bigg|   \,dr\,.
\end{eqnarray*}
Following the outline of the proof of Lemma~1.5~\cite{Y2017} we set   $  r=ye^{b}$ and  obtain 
\begin{eqnarray*} 
&  &
\| \Phi  (x,t) \| _{B^{s',q}_{p'}} \\
 & \lesssim  &
   e^{\frac{n}{2}t}e^{ -\Re Mt }\int_{ 0}^{t}   e^{-( \frac{n}{2}-\Re M )b}    e^{ab}\|f(x,b)\|_{B^{s,q}_{p}} \,db   \nonumber \\
 &  &
 \times 
  \int_{ 0}^{ e^{t-b}- 1}  y^{a}      \Big((e^{t-b }+1)^2 - y^2\Big)^{-\frac{1}{2}+\Re M    }\Bigg| F\Big(\frac{1}{2}-M   ,\frac{1}{2}-M  ;1; 
\frac{ ( e^{t-b}-1)^2 -y^2 }{( e^{t-b}+1)^2 -y^2 } \Big)  \Bigg|   \,dy \nonumber \,.
\end{eqnarray*}
In order to estimate the integral
\[ 
I_{a,M}(z):=\int_{ 0}^{ z- 1} \, y^{a}   \Big(( z+1)^2 - y^2\Big)^{-\frac{1}{2}+\Re M    } \left| F\Big(\frac{1}{2}-M   ,\frac{1}{2}-M  ;1; 
\frac{ (z-1)^2 -y^2 }{(z+1)^2 -y^2 } \Big) \right| dy ,  
\]
where $z=e^{t-b} >1$ and $a> -1$  we use 
\[ 
 \left| F\Big(\frac{1}{2}-M   ,\frac{1}{2}-M  ;1; 
\frac{ (z-1)^2 -y^2 }{(z+1)^2 -y^2 } \Big) \right| \leq C_M 
\]
and  estimate the integral
\begin{eqnarray*} 
I_{a,M}(z)
& \leq &
\int_{ 0}^{ z- 1} \, y^{a}   \Big(( z+1)^2 - y^2\Big)^{-\frac{1}{2}+\Re M    }   dy \\
& = &
\frac{1 }{a+1}(z-1)^{a+1} (z+1)^{2 \Re M-1}  F \left(\frac{a+1}{2},\frac{1}{2}-\Re M;\frac{a+3}{2};\frac{(z-1)^2}{(z+1)^2}\right)\\
& \lesssim &
(z-1)^{a+1} (z+1)^{2 \Re M-1} \,.
\end{eqnarray*}
Hence,  
\begin{eqnarray*} 
&  &
\| \Phi  (x,t) \| _{B^{s',q}_{p'}} \\
 & \lesssim  &
   e^{\frac{n}{2}t}e^{ -\Re Mt }\int_{ 0}^{t}   e^{-( \frac{n}{2}-\Re M )b}    e^{ab} ( e^{t-b}-1)^{1+a}   (e^{(t-b)}+1)^{(2\Re M-1 )} \|f(x,b)\|_{B^{s,q}_{p}}  \,db  \\
 & \lesssim  &  
%   e^{\frac{n}{2}t}e^{ -\Re Mt }\int_{ 0}^{t}   e^{-( \frac{n}{2}-\Re M )b}    e^{ab}   e^{t(1+a)}  e^{-b(1+a)} (e^{(t-b)}+1)^{(2\Re M-1 )} \|f(x,b)\|_{B^{s,q}_{p}}  \,db\\
% & \lesssim  &  
   e^{\frac{n}{2}t}e^{ -\Re Mt }e^{t(1+a)}\int_{ 0}^{t}   e^{-( \frac{n}{2}+1-\Re M )b}          (e^{(t-b)}+1)^{(2\Re M-1 )} \|f(x,b)\|_{B^{s,q}_{p}}  \,db\\
 & \lesssim  &  
%   e^{\frac{n}{2}t}e^{ -\Re Mt }e^{t(1+a)}\int_{ 0}^{t}   e^{-( \frac{n}{2}+1-\Re M )b}           e^{(t-b)(2\Re M-1 )}   \|f(x,b)\|_{B^{s,q}_{p}}  \,db\\
% & \lesssim  &
  e^{ t ( \frac{n}{2}+ \Re M +a)} 
\int_{ 0}^{t}   e^{-( \frac{n}{2}+\Re M )b}           \|f(x,b)\|_{B^{s,q}_{p}}  \,db   \,.
 \end{eqnarray*}
Theorem is proved. \hfill $\square$

\section{Integral equation. Global existence in Sobolev \\
spaces}
\label{S3}

\setcounter{equation}{0}
\renewcommand{\theequation}{\thesection.\arabic{equation}}

We are going to apply the Banach fixed-point theorem.  
In order to estimate nonlinear term we use the   Lipschitz condition (${\mathcal L}$).
Evidently, the Condition ($\mathcal L$) imposes some restrictions on $n$, $\alpha $, $s$. 
First for $A(x,\partial_x)=\Delta  $ we consider the integral equation  (\ref{5.1}),
\begin{eqnarray*} 
\Phi (x,t)
 = 
\Phi _0(x,t) + 
G[ e^{-\Gamma \cdot }F(\cdot ,\Phi ) ] (x,t)\,,      
\end{eqnarray*}
where the function $\Phi _0\in C([0,\infty);H_{(s)} ({\mathbb R}^n))$ is given. 
Every solution to   the Cauchy problem  (\ref{0NWE})-(\ref{CD0.11}) 
 solves also the  integral equation (\ref{5.1}) with some function $\Phi _0=\Phi _0 (x,t)$ 
  which is   a 
solution to the Cauchy problem for the linear equation  (\ref{1.7}).

The operator $G$ and the structure of the nonlinear term determine the solvability of the integral equation (\ref{5.1}). 
For the operator $G $ generated by the linear part of the equation (\ref{0.1}) with $m^2<0 $ the global solvability 
of the integral 
equation (\ref{5.1}) was studied in \cite{yagdjian_DCDS}. For the case of  $m^2<0 $ and the nonlinearity $F(\Phi) = c  |\Phi  |^{\alpha +1} $, $c\not=0$,  the results of \cite{yagdjian_DCDS}
imply the nonexistence of the global solution even for arbitrary small   function $\Phi _0 (x,0) $ under some conditions on $n$  and  $\alpha $.
\begin{theorem}
\label{TIE} 
Assume that  $F(x,\Phi )$  is Lipschitz continuous in the  space $H_{(s)} ({\mathbb R}^n)$,   $F(x,0)\equiv 0$, and  $\alpha >0 $. 
  Suppose that  $\Re M >0  $ and 
one of the following conditions is satisfied
\begin{eqnarray*} 
& (i_0)  &
\quad  \frac{n}{2}+ \Re M   + \gamma (\alpha +1)+\Gamma >0\,,\quad   \frac{n}{2}+  \Re M   + \gamma  \leq  0\,,\\
& (ii_0) &
 \quad  \frac{n}{2}+\Re M  + \gamma (\alpha +1)+\Gamma = 0\,,\quad \frac{n}{2}+ \Re M  + \gamma  < 0\,,\\
& (iii_0) &
 \quad  \frac{n}{2}+\Re M  + \gamma (\alpha +1)+\Gamma < 0\,,\quad  \gamma  \alpha +\Gamma  \geq   0    \,.
\end{eqnarray*}
  Then for every given function $ \Phi _0(x ,t) \in X({\varepsilon ,H_{(s)} ({\mathbb R}^n),\gamma }) $  such that  
\[
\sup_{t \in [0,\infty)}   e^{\gamma  t}  \|\Phi _0(\cdot ,t) \|_{H_{(s)}({\mathbb R}^n)}  < \varepsilon\,,  
\]
and for sufficiently small  $\varepsilon $,   \, the integral equation (\ref{5.1}) has a unique solution \, $ \Phi  (x ,t) \in X({2\varepsilon ,H_{(s)}({\mathbb R}^n),\gamma })  $, that is  
\[  
\sup_{t \in [0,\infty)}    e^{\gamma  t}  \|\Phi  (\cdot ,t) \|_{H_{(s)} ({\mathbb R}^n)}  < 2\varepsilon \,.
\]
\end{theorem}
\begin{theorem}
\label{TIEb}  Assume that  $F(x,\Phi )$  is Lipschitz continuous in the  space $H_{(s)} ({\mathbb R}^n)$,   $F(x,0)\equiv 0$, and  $\alpha >0 $. 
  Suppose that  
one of the following conditions is satisfied
\begin{eqnarray*} 
& (iv_0) &
 \quad  \frac{n}{2}+\Re M  + \gamma (\alpha +1)+\Gamma > 0\,,\quad \frac{n}{2}+ \Re M  + \gamma  >  0 \,,\\
& (v_0) &
 \quad  \frac{n}{2}+\Re M  + \gamma (\alpha +1)+\Gamma = 0\,,\quad \frac{n}{2}+ \Re M + \gamma \geq  0\,,\\
& (vi_0)  &
\quad  \frac{n}{2}+\Re M  + \gamma (\alpha +1)+\Gamma <0\,,\quad  \gamma  \alpha +\Gamma  <   0\,.
\end{eqnarray*}
If $\Re M  > 0$, then for  the function $ \Phi _0(x ,t) \in X({\varepsilon ,H_{(s)}  ({\mathbb R}^n),\gamma }) $     
the  unique solution \, $ \Phi  (x ,t)   $ of the integral equation (\ref{5.1}) has   
the lifespan $T_{ls}$ that can be estimated from below  
\[  
  T_{ls}    
   \geq   {\cal I} \left(C(M,n,\alpha ,\gamma, \Gamma  ) \left(  \max_{\tau  \in [0,\infty)} e^{\gamma  \tau } \|\Phi _0(\cdot ,\tau ) \|_{ H_{(s)} ({\mathbb R}^n) } \right) ^{-\alpha }\right)   
\] 
for sufficiently small $ \max_{\tau  \in [0,\infty)} e^{\gamma  \tau } \|\Phi _0(\cdot ,\tau ) \|_{ H_{(s)} ({\mathbb R}^n) }  $ 
 with some number $C(M,n,\alpha ,\gamma, \Gamma  )>0 $.
\end{theorem}
We need the following elementary lemma. % that will be used with ${\mathcal M}:= {1/2} $ and ${\mathcal M}:=\Re M =\Re (\frac{n^2}{4}-m^2)^{1/2} $.   
\begin{lemma}
\label{L3.2} 
For ${\mathcal M} \in  [0,\infty)$ and $ \alpha >0$   the inequality 
\begin{eqnarray}
\label{Int2}  
&  & e^{ t ( \frac{n}{2}+ {\cal M} + \gamma )} 
\int_{ 0}^{t}   e^{-( \frac{n}{2}+{\cal M} + \gamma (\alpha +1)+\Gamma )b}\,db \leq const \quad \mbox{  for all} \quad t \in [0,\infty)  
\end{eqnarray}  
holds true  
 for all   $\gamma  $ and $\Gamma  $ such that one of these conditions is satisfied
\begin{eqnarray*} 
& (i_{\cal M})  &
\quad  \frac{n}{2}+{\cal M} + \gamma (\alpha +1)+\Gamma >0\,,\quad   \frac{n}{2}+ {\cal M} + \gamma  \leq  0\,,\\
& (ii_{\cal M}) &
 \quad  \frac{n}{2}+{\cal M} + \gamma (\alpha +1)+\Gamma = 0\,,\quad \frac{n}{2}+ {\cal M} + \gamma  < 0\,,\\
& (iii_{\cal M}) &
 \quad  \frac{n}{2}+{\cal M} + \gamma (\alpha +1)+\Gamma < 0\,,\quad  \gamma  \alpha +\Gamma  \geq   0    \,.
\end{eqnarray*}
If one of the following conditions
\begin{eqnarray*} 
& (iv_{\cal M})  &
\quad  \frac{n}{2}+{\cal M} + \gamma (\alpha +1)+\Gamma > 0\,,\quad   \frac{n}{2}+ {\cal M} + \gamma > 0\,,\\
& (v_{\cal M}) &
 \quad  \frac{n}{2}+{\cal M} + \gamma (\alpha +1)+\Gamma = 0\,,\quad \frac{n}{2}+ {\cal M} + \gamma \geq  0\,,\\
& (vi_{\cal M}) &
 \quad  \frac{n}{2}+{\cal M} + \gamma (\alpha +1)+\Gamma < 0\,,\quad  \gamma  \alpha +\Gamma  <   0\,,
\end{eqnarray*}
is fulfilled, then the function 
\begin{eqnarray*} 
  I_{\cal M}(t):=  e^{ t  ( \frac{n}{2}+ {\cal M} + \gamma )} 
\int_{ 0}^{t }   e^{-( \frac{n}{2}+{\cal M} + \gamma (\alpha +1)+\Gamma )b}\,db  
\end{eqnarray*} 
is monotonic and unbounded, 
$  
\lim_{t \to \infty} I_{\cal M}(t) =  \infty  
$.
\end{lemma}
\medskip

\noindent
{\bf Proof of Theorem~\ref{TIE}.} 
 Consider the mapping
\begin{eqnarray*} 
 S[\Phi ] (x,t)
& := &
\Phi _0(x,t) + 
G[e^{-\Gamma \cdot }  F(\cdot , \Phi  )] (x,t)    \,. 
\end{eqnarray*} 
We are going to prove that $S$ maps $X({R,H_{(s)} ({\mathbb R}^n) ,\gamma })$ into itself and that $S$ is a contraction, provided that   $\varepsilon  $ and $R$ are sufficiently small. 
Theorem~\ref{T11.1}    implies
\[
  \|   S[\Phi ] (\cdot  ,t) \|_{ H_{(s)}({\mathbb R}^n)  }  
\leq 
   \|\Phi _0(\cdot ,t) \|_{H_{(s)}({\mathbb R}^n)  }  + C_M   e^{ t ( \frac{n}{2}+  \Re M   
 )} 
\int_{ 0}^{t}   e^{-( \frac{n}{2}+ \Re M  +\Gamma)b}    
 \| \Phi  (\cdot ,b ) \| _{H_{(s)}({\mathbb R}^n)}   ^{\alpha  +1}  \,db\,. 
\]
Taking into account the Condition (${\mathcal L}$)   
we arrive at 
\begin{eqnarray*} 
 e^{\gamma  t} \|   S[\Phi ] (\cdot  ,t) \|_{ H_{(s)}({\mathbb R}^n)  } 
&  \leq   &
  e^{\gamma  t}    \|\Phi _0(\cdot ,t) \|_{ H_{(s)}({\mathbb R}^n)  }  +C_M  \left( \sup_{\tau  \in [0,\infty)} e^{\gamma  \tau  }     \| \Phi  (\cdot ,\tau  ) \| _{ H_{(s)}({\mathbb R}^n)  }    \right)^{\alpha  +1}  \\
&  &
\times   e^{ t ( \frac{n}{2}+  \Re M   + \gamma )} 
\int_{ 0}^{t}   e^{-( \frac{n}{2}+ \Re M  + \gamma (\alpha +1)+\Gamma )b}\,db  \,.
\end{eqnarray*} 
Then, for given $\alpha  $, $\Gamma  $, and    $\gamma \in {\mathbb R} $   according to Lemma~\ref{L3.2}, we have  
\[
\sup_{t  \in [0,\infty)}e^{\gamma t}\|   S[\Phi ] (x ,t) \|_{ H_{(s)}({\mathbb R}^n)  } 
   \leq   
  \sup_{t  \in [0,\infty)}e^{\gamma t}\|\Phi _0(\cdot ,t) \|_{ H_{(s)} ({\mathbb R}^n)  }  +  C      
 \Big( \sup_{t  \in [0,\infty)} e^{\gamma   t  }     \| \Phi  (\cdot ,\tau  ) \| _{ H_{(s)}({\mathbb R}^n)}   \Big)^{\alpha  +1} . 
\]
 Thus, the last inequality proves that the operator $S$ maps $X({R, H_{(s)}({\mathbb R}^n)  ,\gamma})$ into itself if $\varepsilon  $ and $R$ are sufficiently small, namely, if
$\varepsilon  +C R^{\alpha +1} < R $.

It remains to prove that $S$ is a contraction mapping.
As a matter of fact, we just  use the estimate (\ref{calM})     in order to obtain   the contraction property  
\begin{equation}
\label{4.17}
 e^{\gamma t} \|S[\Phi ](\cdot,t) -  S[\Psi  ](\cdot,t) \|_{H_{(s)} ({\mathbb R}^n) } 
 \leq  
CR(t) ^{\alpha } d(\Phi ,\Psi )\,,  
\end{equation}
where $\displaystyle   R(t):= \max\{ \sup_{0\leq \tau \leq t } e^{\gamma \tau  }  \| \Phi  (\cdot ,\tau ) \| _{H_{(s)} ({\mathbb R}^n)  } , \sup_{0\leq \tau \leq t } e^{\gamma \tau  }  \| \Psi  (\cdot ,\tau ) \| _{H_{(s)} ({\mathbb R}^n) }\} \leq R$.

Indeed, the inequality (\ref{Int2}) with ${\mathcal M}=\Re M $ 
 holds true  for $\alpha  $ and $\gamma  $ satisfying conditions of the theorem. 
Together with
\[
e^{\gamma t} \|S[\Phi ](\cdot,t) -  S[\Psi  ](\cdot,t) \|_{H_{(s)} ({\mathbb R}^n) } \\
  \leq  
C _M   d(\Phi ,\Psi ) R(t)^\alpha e^{ t ( \frac{n}{2}+ \Re M  + \gamma )} 
\int_{ 0}^{t}   e^{-( \frac{n}{2}+\Re M  + \gamma (\alpha +1)+ \Gamma )b}\,db  
\] 
it leads to (\ref{4.17}). 
Next we choose $\varepsilon $ and $ R$ sufficiently small.  Banach's fixed point 
theorem completes the proof.\hfill $\square$
 \smallskip

\noindent
{\bf Proof of Theorem~\ref{TIEb}.} We have   
\begin{eqnarray*}  
e^{\gamma t}\| \Phi  (\cdot  ,t) \|_{  H_{(s)}({\mathbb R}^n)  }  
&  \leq   &
 e^{\gamma t} \|\Phi _0(\cdot ,t) \|_{  H_{(s)}({\mathbb R}^n)  }   
+ C_M I(t) \left( \sup_{\tau  \in [0,\infty)} e^{\gamma  \tau  }     \| \Phi  (\cdot ,\tau  ) \| _{  H_{(s)}({\mathbb R}^n)  }    \right)^{\alpha  +1}  \,. 
\end{eqnarray*}
Set 
 \[
 T_\varepsilon
  :=  
\inf \{ T\,:\,\max_{\tau  \in [0,T]}e^{\gamma  \tau }  \|    \Phi   (x ,\tau ) \|_{ H_{(s)} ({\mathbb R}^n)  } \geq 2\varepsilon \}\,, 
\quad  \varepsilon 
   := 
   \max_{\tau  \in [0,\infty)} e^{\gamma  \tau } \|\Phi _0(\cdot ,\tau ) \|_{ H_{(s)} ({\mathbb R}^n) }  \,.
 \]
 Then
\[  
2\varepsilon    \leq   
\varepsilon  +  C_{M,n,\alpha }       
 \varepsilon  ^{\alpha  +1}  
I_{M}(T_\varepsilon)
 \] 
 implies
\[  
I_{M}(T_\varepsilon) \geq C_{M,n,\alpha }^{-1} \epsilon^{-\alpha }  \,. 
\] 
If $\cal I$ is a  function inverse to $I=I(t)$, then  
\[  
  T_{ls}    
   \geq   {\cal I}_{M} \left(C_{M,n,\alpha }^{-1} \left(  \max_{\tau  \in [0,\infty)} e^{\gamma  \tau } \|\Phi _0(\cdot ,\tau ) \|_{ H_{(s)} ({\mathbb R}^n) } \right) ^{-\alpha }\right)  \,. 
\] 
for sufficiently small $ \max_{\tau  \in [0,\infty)} e^{\gamma  \tau } \|\Phi _0(\cdot ,\tau ) \|_{ H_{(s)} ({\mathbb R}^n) }  $. 
  Theorem is proved. \hfill $\square$

\section{The Cauchy problem. Global existence of small data solutions}
\label{S4}
\setcounter{equation}{0}
\renewcommand{\theequation}{\thesection.\arabic{equation}}

 {\bf Proof of Theorem~\ref{T0.2}. } 
For the function   $\Phi_0 $, that is, for the solution of the Cauchy problem (\ref{1.7})  and for   $n \geq 2$, according to Theorem~\ref{T13.2}  
we have with $a= 0 $ the estimate
\begin{eqnarray*}  
&  &
e^{\gamma t} \| \Phi_0  (x,t) \| _{H_{(s)}}   \\
&   \lesssim  &  
\|\varphi _0 \|_{H_{(s)}}  e^{(\frac{n}{2}+ \Re M+\gamma )t}    
\cases{ 1 \quad \mbox{\rm if } \quad   \Re \, M>1/2 \cr  
t^{\sgn|\Im M|}+ e^{(\frac{1}{2}-\Re M)t  }   \quad \mbox{\rm if } \quad \Re \,M  \leq  1/2 } \nonumber \\ 
&  & 
\,+ \|\varphi _1 \|_{H_{(s)}} e^{(\frac{n}{2} +\Re M+\gamma )t}
  \\   
&   \lesssim  &  
 e^{(\frac{n}{2}+ \Re M+\gamma )t} \Bigg( \|\varphi _1 \|_{H_{(s)}} + \|\varphi _0 \|_{H_{(s)}}    
\cases{ 1 \quad \mbox{\rm if } \quad   \Re \, M>1/2 \cr  
t^{\sgn|\Im M|}+ e^{(\frac{1}{2}-\Re M)t  }   \quad \mbox{\rm if } \quad \Re \,M  \leq  1/2 } \Bigg)
\end{eqnarray*}  
for all   $t \in (1,\infty)$. The factors of the norms of the initial data functions   of the last inequality are bounded provided that
\[
 \frac{n}{2}+ \max\{ \frac{1}{2}, \,\Re M\}+\gamma \leq 0, \quad \Re \,M  <  1/2 \quad \mbox{\rm or} \quad \Re \,M  >  1/2 \quad \mbox{\rm or} \quad M= \Re \,M  =  1/2    \,.
\]  
 Thus, 
  we have $\Phi_0 \in X(R, H_{(s)} ({\mathbb R}^n),\gamma ) $. Consequently, according to Theorem~\ref{TIE} 
the function $\Phi $ belongs to the space  $X(R, H_{(s)} ({\mathbb R}^n),\gamma ) $, where the operator $S $ is a contraction. 
Theorem is proved. \hfill $\square$

\noindent
\noindent{\bf Proof of Theorem~\ref{T0.2b}. }     According to Theorem~\ref{T11.1} 
we have the estimate 
\begin{eqnarray*}   
 e^{\gamma  t} \|   S[\Phi ] (\cdot  ,t) \|_{  H_{(s)} ({\mathbb R}^n)  }   
&  \leq   &
  e^{\gamma  t}    \|\Phi _0(\cdot ,t) \|_{  H_{(s)} ({\mathbb R}^n)  }  +C_M  \left( \sup_{\tau  \in [0,\infty)} e^{\gamma  \tau  }     \| \Phi  (\cdot ,\tau  ) \| _{ H_{(s)} ({\mathbb R}^n)  }    \right)^{\alpha  +1}  \\
&  &
\times   e^{ t ( \frac{n}{2}+ \Re M + \gamma )} 
\int_{ 0}^{t}   e^{-( \frac{n}{2}+\Re M + \gamma (\alpha +1)+\Gamma )b}\,db  \,.
\end{eqnarray*}   Due to Corollary~\ref{C1.4} we have $\Phi_0 \in X(R, H_{(s)} ({\mathbb R}^n),\gamma ) $.
Thus,  
\begin{eqnarray*}   
&  &
 e^{\gamma  t} \|   S[\Phi ] (\cdot  ,t) \|_{ H_{(s)}({\mathbb R}^n)  } \\  
&  \leq   &
C e^{(\frac{n}{2}+ \max\{ \frac{1}{2}, \,\Re M\}+\gamma )t}  \left(  \|\varphi _0 \|_{L^{p}}  +\|\varphi _1 \|_{H_{(s)}} \right) 
+C_M  \left( \sup_{\tau  \in [0,\infty)} e^{\gamma  \tau  }     \| \Phi  (\cdot ,\tau  ) \| _{ H_{(s)}({\mathbb R}^n)  }    \right)^{\alpha  +1}   I(t) \\   
&  \leq   &
C \left(  \|\varphi _0 \|_{L^{p}}  +\|\varphi _1 \|_{H_{(s)}} \right)
+C_M  \left( \sup_{\tau  \in [0,\infty)} e^{\gamma  \tau  }     \| \Phi  (\cdot ,\tau  ) \| _{ H_{(s)}({\mathbb R}^n)  }    \right)^{\alpha  +1}   I(t) \,.
\end{eqnarray*}
 Set 
 \[ 
 T_\varepsilon
   :=  
\inf \{ T\,:\,\max_{\tau  \in [0,T]}e^{\gamma  \tau }  \|    \Phi   (x ,\tau ) \|_{H_{(s)}({\mathbb R}^n)  } \geq 2\varepsilon \}\,,  
\quad  \varepsilon 
   :=  
   C( \|\varphi _0 \|_{H_{(s)}}  +\|\varphi _1 \|_{H_{(s)}} )\,.
 \]
 Then 
 \[  
2\varepsilon    \leq   
\varepsilon  +  C_{M,n,\alpha }       
 \varepsilon  ^{\alpha  +1}  I( T_\varepsilon)
 \] 
 implies the statement  of the theorem. 
 Thus, the theorem is proved.  \hfill $\square$

\end{document}